\documentclass[journal]{new-aiaa}
\usepackage[utf8]{inputenc}
\usepackage{graphicx}
\usepackage{amsmath,bm}
\usepackage[version=4]{mhchem}
\usepackage{longtable,tabularx}
\usepackage{subfigure}
\usepackage{float}
\usepackage{siunitx}
\usepackage{boldline}
\usepackage{multirow}
\usepackage{algorithm,algcompatible}
\usepackage{algpseudocode}
\usepackage[table]{xcolor}
\usepackage{booktabs}
\usepackage{dutchcal}
\usepackage{setspace}
\usepackage{bm}
\usepackage{array}
\usepackage{dcolumn}
\usepackage{soul}
\usepackage{xcolor}
\usepackage{ulem}
\soulregister\cite7
\soulregister\citep7
\soulregister\citet7
\soulregister\ref7
\soulregister\eqref7
\definecolor{bananamania}{rgb}{0.98, 0.91, 0.71}
\sethlcolor{bananamania}

\newcolumntype{d}[1]{D{.}{.}{#1}}
\DeclareMathAlphabet\mathbfcal{OMS}{cmsy}{b}{n}
\newcolumntype{P}[1]{>{\centering\arraybackslash}p{#1}}

\AtBeginEnvironment{algorithmic}{\onehalfspacing}
\algnewcommand\Given{\item[\textbf{Given:}]}%
\algnewcommand\Initialize{\item[\textbf{Initialize:}]}%
\DeclareMathAlphabet{\mathcal}{OMS}{cmsy}{m}{n}
\newcommand*\dif{\mathop{}\!\mathrm{d}}

\title{Improved Stability-Based Transition Transport Model for Airships Incorporating Wall Heating Effects}

\author{Yayun Shi\footnote{Associate Professor, State Key Laboratory for Strength and Vibration of Mechanical Structures, School of Aerospace Engineering.}}
\affil{Xi'an Jiaotong University, Xi'an,  Shaanxi, 710049, China}
\author{Qiyun Wang\footnote{Master, National Key Laboratory of Aircraft Configuration Design, School of Aeronautics.}}
\affil{Northwestern Polytechnical University, Xi'an, Shaanxi, 710072, China}
\author{Xiaosong Lan\footnote{Master,  State Key Laboratory for Strength and Vibration of Mechanical Structures, School of Aerospace Engineering.}} 
\affil{Xi'an Jiaotong University, Xi'an,  Shaanxi, 710049, China}
\author{Bo Wang\footnote{Research Fellow,  Qingdao Institute of Aeronautical Technology.}} 
\affil{Qingdao Institute of Aeronautical Technology, Qingdao, Shandong, 266400, China}
\author{Tihao Yang\footnote{Associate Research Fellow, National Key Laboratory of Aircraft Configuration Design, School of Aeronautics.}}
\affil{Northwestern Polytechnical University, Xi'an, Shaanxi, 710072, China}
\author{Yifu Chen\footnote{PhD candidate, Aerodynamics Group, Faculty of Aerospace Engineering. Email: y.chen-20@tudelft.nl (Corresponding Author)}}
\affil{Delft University of Technology, Kluyverweg 1, 2629HS Delft, The Netherlands}

\begin{document}
\maketitle

\begin{abstract}
Laminar drag reduction is a critical technology for enhancing the endurance and station-keeping capabilities of airship platforms.
However, existing transport-based transition models fail to account for the premature transition induced by wall heating, a limitation that significantly hinders the robust engineering application of laminar-flow technology in realistic thermal environments.
To address this deficiency, this study first develops stability-based correction for transition modeling that explicitly incorporates wall-to-freestream temperature ratios.
Leveraging the Falkner--Skan--Cooke (FSC) equations and linear stability theory (LST) with the $e^N$ method, we derive physics-based correlations for the transition criteria as functions of the temperature ratio, pressure gradient, and turbulence intensity.
These corrections are integrated into a simplified stability-based transition transport model proposed by \citet{franccois2023simplified} and validated against the classic Schubauer and Klebanoff flat-plate experiments, demonstrating accurate prediction of transition locations under adiabatic, heated, and cooled conditions.
Crucially, wind-tunnel experiments on a heated airship model show that wall-heating sensitivity is strongly influenced by local pressure-gradient variations, which is due to Reynolds-number-driven transition-location shifts.
The proposed model successfully reproduces the experimentally observed transition advancement caused by wall heating.
This framework, covering both heating and cooling regimes, provides a capability to support future laminar-flow control technologies based on wall-temperature modulation.
%  transition criteria and localized model parameters that incorporate the effects of the wall-to-freestream temperature difference, based on the FSC equation, the linear stability equation, and the $\bm{e^N}$ method.
% We then implement the proposed transition criteria and localized ratio parameters into an intermittency-equation transition model and validate it against the Schubauer and Klebanoff flat-plate case. The results demonstrate that the improved transition model can accurately capture temperature-difference effects.
% Importantly, we conduct transition experiments on an airship configuration, including both an adiabatic wall and a heated wall.
% Further comparison with airship wind-tunnel measurements confirms that our improved model can capture the transition-advancement phenomenon induced by wall heating.
% Additionally, we consider both wall heating and wall cooling within our modeling framework, enabling accurate transition prediction to support future laminar drag-reduction technologies based on controlled wall-temperature variation.

\end{abstract}

\section{Introduction}
% This manuscript summarizes an approach to reduce design-space nonlinearity and geometric distortion for adjoint-driven laminar airfoil optimization. The aim is a more robust, automated design workflow with limited manual tuning.
% 飞艇具有重要的应用价值，尤其是在军事和民用领域\cite{guo2020recent}。与传统的固定翼飞机相比，飞艇具有更高的燃油效率和更低的运营成本\cite{bryson2008design}。飞艇得益于其流线设计，使得层流减阻技术成为其重要的提升气动性能的关键因素。但是，飞艇在白天由于太阳能电池受到太阳辐射等外部因素的影响，导致其表面温度远高于环境温度，从而影响层流-湍流的转捩过程\cite{li2019numerical}，进一步影响层流减阻效果。因此，准确预测壁面升温条件下飞艇表面的层流-湍流转捩位置对于提升飞艇的性能至关重要。

\begin{figure}[h!]
  \centering
  \includegraphics[width=0.7\textwidth]{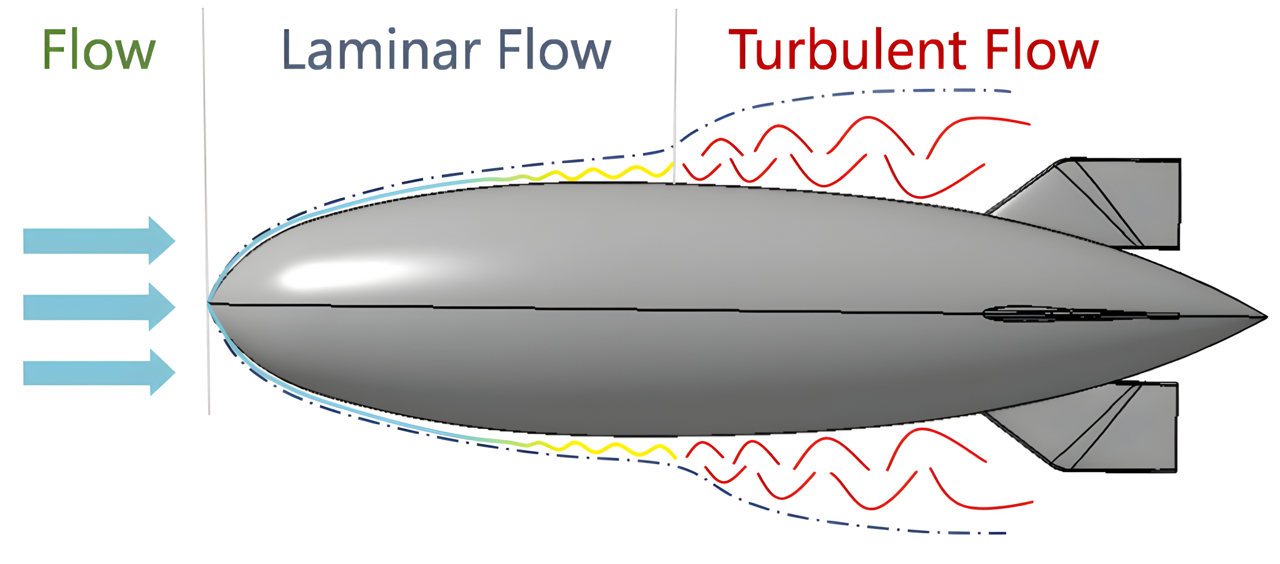}
  \caption{The schematic diagram of laminar flow drag reduction on an airship.}
  \label{f:Airship}
\end{figure}

Stratospheric and high-altitude airships hold significant promise for both military and civil applications due to their exceptional endurance and station-keeping capabilities\cite{ilieva2012critical, liao2009review}.
Relative to conventional fixed-wing aircraft, these platforms offer superior energy efficiency and reduced operational costs\cite{zuo2022survey, wu2015thermal}.
To maximize station-keeping capabilities, achieving superior aerodynamic efficiency is essential.
Given their streamlined configuration (Fig.~\ref{f:Airship}), implementing laminar-flow drag-reduction techniques becomes a primary strategy for realizing this performance goal.
However, the operational environment introduces complex adverse factors; specifically, during daytime, solar usage and direct radiation can elevate the surface temperature significantly above ambient levels\cite{celep2022effect, lv2025numerical, kral1994direct}. This wall heating fundamentally alters boundary-layer stability, causing premature laminar–turbulent transition (LTT) and severely compromising drag-reduction benefits. Consequently, accurate prediction of LTT location under realistic thermal conditions is a prerequisite for robust airship design.

% 包括壁面升温在内，壁面冷却也会对转捩位置产生影响\cite{}。针对不同研究对象，壁面升温和降温对转捩的影响不同。针对绕水流动，升温会抑制扰动，延迟转捩。对于高超声速第一模态扰动，升温会提前转捩，但针对第二模态，降温会导致转捩提前。对于超声速流动，升温会。。。而对于低速、亚声速流动流动中的流向转捩和横流转捩，升温会将转捩提前，即壁面冷却可以延迟转捩，这也是本文研究对象的流动速域。
% Both wall heating and cooling can influence the transition location\cite{celep2022effect, lv2025numerical}. The temperature gradients in the boundary layer on transition differ depending on the flow regime and disturbance type. For a surface immersed in water flow, wall heating tends to suppress disturbances and delay LLT transition\cite{lv2023numerical, lv2025numerical}.
% In hypersonic flows, uniform and localized wall heating/cooling affect transition differently for first-mode and second-mode disturbances\cite{zhao2018numerical,liu2024roughness,xu2022wall}.
% In supersonic flows, wall heating triggers transition earlier, while cooling delays it~\cite{celep2022effect}.
% However, in the coupling of temperature and roughness elements, wall heating suppresses the disturbance growth induced by the roughness elements~\cite{liu2024roughness}.
% In low-speed and subsonic regimes, wall heating amplifies boundary-layer disturbances whereas wall cooling delays transition, and this study centers on that behavior~\cite{manuilovich2014heat,lopez2024impact}.

The modulation of transition by wall temperature is highly dependent on the flow regime and the dominant instability mechanism\cite{celep2022effect, lv2025numerical}.
In hydrodynamic flows (water), wall heating typically stabilizes Tollmien-Schlichting (TS) waves, delaying transition\cite{lv2023numerical, lv2025numerical}.
In high-speed aerodynamics, the effects are mode-dependent: for hypersonic flows, wall heating promotes first-mode instabilities but stabilizes second-mode disturbances\cite{zhao2018numerical,liu2024roughness,xu2022wall}, while in supersonic flows, heating generally triggers earlier transition\cite{celep2022effect}.
Conversely, in the subsonic (low subsonic and transonic) regimes relevant to airships, wall heating has a distinct destabilizing effect, amplifying disturbances and advancing transition, whereas wall cooling acts as a stabilizer\cite{manuilovich2014heat,lopez2024impact}.
This study focuses specifically on quantifying and modeling this destabilizing behavior in the subsonic regime.

% 对于不同的研究对象，从解释机理方面，主要是以高精度数值模拟和稳定性理论为主。基于稳定性理论的e^N方法也被用于水下航行器的工程转捩预测。
% 然而，LST方法计算复杂且难以直接应用于工业级三维流场中的转捩预测，尤其是针对飞艇这类构型。相比之下，如Langtry-Menter模型\cite{langtry2009correlation}由于其完全依赖当地变量适用于CFD并行与非机构网格，已被广泛应用于工业界。
% 在高超声速领域，Kovalev在Langtry-Menter模型上，将温度比作为修正因子引入到转捩准则中，并研究边界层温度梯度对高超声速飞行器转捩位置的影响\cite{kovalev2014transition}。
% 徐等人\cite{shen2025novel}针对水下航行器，结合FSC方程和LST方程，标定了受温度影响的单方程关键转捩参数。该单方程从Langtry-Menter两方程演化而来。修正后的模型能够准确的预测带有边界层温度梯度影响的水下航行器的转捩位置。
Across different flow regimes, the underlying transition mechanisms have mainly been investigated using direct numerical simulation (DNS) and stability-based approaches.
\citet{lv2025numerical} use the DNS to solve the laminar boundary layer equations with costant wall temperature conditions. With this base flow, the LST analysis is performed to study the effects of wall temperature on the disturbance growth and transition process.
The simulation results agree very well with the experimental data.
\citet{fedorov2015high} employed DNS and LST approches to investigate the hypersonic boundary-layer stability on a cone with localized wall heating/cooling, and the LST computation of the second-mode amplification agrees well with the DNS solutions.
Similar approaches have also been applied in supersonic and subsonic flows~\cite{poulain2024adjoint,thummar2024stability}, and demonstated that the stability analysis methods can accurately caputure the heating/cooling effects.

% 想介绍下LST的一些改进以及验证

However, stability-based transition prediction is computationally complex and difficult to apply directly to industrial three-dimensional configurations with complex geometries, such as the airship considered in this work.
Meanwhile, the optimization of such configurations often requires numerous iterations, making stability-based methods impractical for routine use~\cite{Shi2020a,Shi2021a,shi2023complex,chen2023adjoint}.
In contrast, correlation-based transport models such as the Langtry--Menter model\cite{Langtry2009}, which rely solely on local flow variables and are well suited for parallel CFD and unstructured grids, have been widely adopted in industry.
Similar transport-based transition models have also been extensively developed~\cite{Coder2014a,krimmelbein2021validation,zhou2019application}.
Consequently, a growing body of research aims to bridge this gap by deriving physics-informed corrections from high-fidelity datasets—generated via LST—and integrating them into transport models. This hybrid strategy allows for the accurate accounting of specific flow physics, such as crossflow or surface roughness, while retaining computational efficiency\cite{gosin2025assessment,krumbein2022transport,liu2020two,Choi2015,grabe2018transport,xu2019fully,qin2020predicting}.
\citet{franccois2023simplified} recently proposed a simplified one-equation transition model based on the Langtry--Menter framework, which incorporates stability-based criteria for key transition parameters derived from LST analysis.
% 类似的输运模型也得到了广泛发展。

% 基于高精度数值模拟和稳定性理论，可以对输运模型关键转捩准则进行补全与修正，从而考虑不同因素的影响~\cite{}。
% 针对壁面升温降温的影响，在高超声速和水下流动中，已经有了相关研究。

With regard to the effects of wall heating or cooling, related studies have already been carried out in hypersonic and underwater flows.
In the hypersonic regime, Kovalev\cite{kovalev2015modeling} incorporated the wall-temperature ratio as a correction factor into the Langtry--Menter transition criterion and investigated the influence of wall-to-freestream temperature difference on the transition location of hypersonic vehicles.
For underwater vehicles, \citet{shen2025novel} combined the FSC equations with LST to calibrate key temperature-dependent transition parameters in the one equation model~\cite{menter2015one,franccois2023simplified} derived from the original two-equation Langtry--Menter formulation. The corrected model is able to accurately predict the transition location of underwater vehicles in the presence of boundary-layer temperature gradients.

For the airship configuration considered in this work, transport-based transition models are required to accelerate and simplify the aerodynamic shape optimization process. 
To the best of the authors’ knowledge, no studies have yet incorporated the effects of wall heating into transport-based transition models for the subsonic airflows.
As a result, when designing laminar airship configurations using such models, the influence of wall heating cannot be accounted for in robust laminar-flow design. 
This, in turn, leads to overly optimistic estimates of laminar drag reduction, because the detrimental impact of wall heating is not accounted for the outset of the design process, preventing truly robust solutions from being obtained.

To address this deficiency, this work introduces a targeted correction to the simplified stability-based transport transition model proposed by \citet{franccois2023simplified}, extending its validity to include generalized wall heating/cooling effects.
Our methodology integrates novel transition criteria and local parameter correlations, derived via linear stability analysis (LST) and the $e^N$ method, into this efficient transport-based framework.
This physics-informed approach specifically captures the wall temperature modulation (both wall heating and wall cooling) of Tollmien-Schlichting (TS) waves—the primary instability mechanism governing transition on streamlined airship hulls at cruise angles of attack ($0^{\circ}$).
Further, we validate the improved model through a dedicated wind-tunnel experiment on a heated airship configuration.
The experimental results provide direct confirmation that the proposed model accurately reproduces the significant upstream shift of the transition front induced by wall heating.
As a result, this validated framework provides a robust engineering capability for the aerodynamic design and optimization of future laminar-flow platforms incorporating wall-temperature effects.

% To address this gap, this paper presents a Langtry--Menter transition-model correction that covers both wall heating and wall cooling in low-speed and subsonic flows, extending beyond the heated airship case to support transition prediction for future vehicles operating in the same regime.
% The TS-wave-induced transition is the dominant mechanism for the examined airship geometry because these vehicles usually cruise close to $0^{\circ}$ angle of attack.
% One of our additional contributions is a heated-airship transition experiment, in which the effect of wall heating on transition was directly observed.
% We then use this experiment to validate the accuracy of the proposed transition model.
% The results demonstate that our proposed correction can accurately capture the influence of wall heating on transition in a real three-dimensional configuration.

The remainder of this paper is structured as follows: Section~\ref{s:modelformulation} presents the theoretical formulation, encompassing the FSC equations, LST analysis, and the proposed corrections to the transition criteria. Section~\ref{s:transportequation} details the implementation of the improved stability-based transport model and evaluates its performance against the conventional LST-based approach for typical cases. Section~\ref{s:Validation} describes the wind-tunnel experiments and validates the improved transport model against experimental measurements from the heated airship configuration. Finally, Section~\ref{s:Conclusions} summarizes the key findings and conclusions.
% 文中要强调，升温对艇随着雷诺数增大，影响会更加明显！

\section{Model Formulation}
\label{s:modelformulation}
\subsection{The Falkner-Skan-Cooke Boundary Layer Equations } % Tollmien Schlichting 
\begin{figure}[h!]
  \centering
  \includegraphics[width=0.5\textwidth]{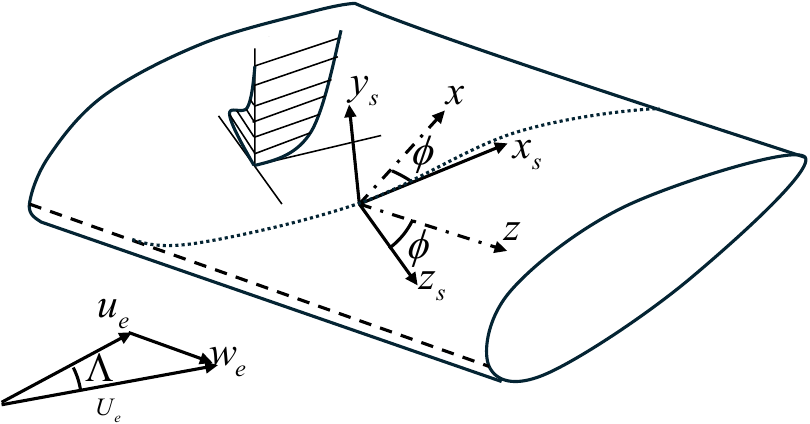}
  \caption{Coordinate system used to formulate the compressible laminar boundary layer equations.}
  \label{f:InfSweptWing}
\end{figure}

Figure~\ref{f:InfSweptWing} illustrates an infinite swept wing along with two coordinate systems.
The first is the wing-attached system, where the $x$-axis aligns with the wing chord and the $z$-axis extends along the spanwise direction.
The second is the streamline coordinate system, in which $x_s$ generally denotes the external potential flow direction.
The angle between these two coordinate systems is denoted by $\phi$.
The three-dimensional compressible laminar boundary layer equations in the wing-attached coordinate system are expressed as follows:
\begin{equation}
  \label{eq:cbl}
  \begin{aligned}
  &\text{Continuity:} \quad \frac{\partial}{\partial x}(\rho u)+\frac{\partial}{\partial y}(\rho v)=0 \\
  &x\text{ momentum:} \quad \rho\left(u \frac{\partial u}{\partial x}+v \frac{\partial u}{\partial y}\right) = \rho_e u_e \frac{\partial u_e}{\partial x} +\frac{\partial}{\partial y}\left(\mu \frac{\partial u}{\partial y}\right) \\
  &z\text{ momentum:} \quad \rho\left(u \frac{\partial w}{\partial x}+v \frac{\partial w}{\partial y}\right) = \frac{\partial}{\partial y}\left(\mu \frac{\partial w}{\partial y}\right) \\
  &\text{Energy:} \quad \rho c_p\left(u \frac{\partial T}{\partial x}+v \frac{\partial T}{\partial y}\right)+u \rho_e u_e \frac{d u_e}{d x}=\frac{\partial}{\partial y}\left(\kappa \frac{\partial T}{\partial y}\right)+\mu\left(\frac{\partial u}{\partial y}\right)^2+\mu\left(\frac{\partial w}{\partial y}\right)^2 \\
  &\text{Gas state:} \quad p = \rho RT
  \end{aligned} ,
\end{equation}
where $u$, $v$, and $w$ denote the velocity components in the $x$, $y$, and $z$ directions, respectively.
$T$ is the temperature, $\rho$ is the density, $\mu$ is the dynamic viscosity, $\kappa$ is the thermal conductivity, $c_p$ is the specific heat at constant pressure, and $R$ is the specific gas constant.
The subscript $e$ indicates the boundary-layer edge.
The no-slip condition is enforced at the wall, combined with either an isothermal or adiabatic thermal condition. The corresponding boundary conditions are given by:
\begin{equation}
  	\label{eq:bcbl}
	\begin{gathered}
		y=0: \quad u=v=w=0, \quad T=T_w \quad \text { or } \quad \frac{\partial T}{\partial y}=0, \\
		y \rightarrow \infty: \quad u \rightarrow u_e, v \rightarrow 0, w \rightarrow w_e, \quad T \rightarrow T_e .
	\end{gathered}
\end{equation}
The subscript of $w$ denotes the wall.

By introducing the similarity variables proposed by Falkner, Skan, and Cooke\cite{stewartson1954further, cooke1950boundary, liu2021compressible}, the governing equations (Eq.\eqref{eq:cbl}) can be rewritten as Eq.\eqref{eq:fsc}: 
\begin{equation}
\label{eq:fsc}
\begin{gathered}
\left(\mathcal{N} f^{\prime \prime}\right)^{\prime}+f f^{\prime \prime}=\frac{2 m}{(m+1)}\left(f^{\prime 2}-\tau\right) \\
\left(\mathcal{N} g^{\prime}\right)^{\prime}+f g^{\prime}=0  \\
\frac{1}{P r}\left(\mathcal{N} \tau^{\prime}\right)^{\prime}+f \tau^{\prime}=0  \\
\mathcal{N}=\tau^{1 / 2}\left(\frac{1+T_s / T_e}{\tau+T_s / T_e}\right) 
\end{gathered},
\end{equation}
known as the compressible Falkner-Skan-Cooke (FSC) equations, where
\begin{equation}
f(\eta)=\frac{u}{u_e}, \quad g(\eta)=\frac{w}{w_e}, \quad \tau(\eta)=\frac{T}{T_e} \quad \text { and } \quad \eta=y \sqrt{\frac{(m+1) u_e}{2 \nu_e x}} .
\end{equation}
The corresponding boundary conditions are:
\begin{equation}
\begin{gathered}
\eta=0: f=f^{\prime}=g=0, \quad \tau=\tau_w \quad \text{(isothermal)} \quad \text { or } \quad \tau^{\prime}=0 \quad \text{(adiabatic)},\\
\eta \rightarrow \infty: f^{\prime}=g=\tau=1 .
\end{gathered}
\end{equation}
Here, $\nu$ denotes the kinematic viscosity, $\Pr$ is the Prandtl number, and $\beta_h =2m/(m+1)$ represents the Hartree pressure-gradient parameter.
The resulting system of ordinary differential equations (ODEs) is solved locally using the Runge--Kutta method.
For the numerical discretization, an equidistant grid is employed in the streamwise direction, while a Chebyshev spectral method is utilized in the wall-normal direction.
Using the solution to the FSC equations, the velocity components projected along the external potential flow direction are expressed as:
\begin{equation}
	\label{eq:velocities_external_flow}
	\begin{aligned}
		& u_s/Q_e = f' \cos^2 \phi + g \sin^2 \phi \\
		& w_s/Q_e = (g -f ') \cos \phi \sin \phi \\
	\end{aligned}.
\end{equation} 
In this study, we focus specifically on the streamwise velocity component, $u_s$.
In Eq.~\eqref{eq:velocities_external_flow}, $Q_e$ denotes the magnitude of the velocity vector at the boundary-layer edge.

\subsection{The Linear Stability Theory Equations and \texorpdfstring{$\boldsymbol{e^N}$}{e\textasciicircum N} Method} % Tomien Schlichting 

The flow state variables are decomposed into the base flow quantities $\bar{\boldsymbol{q}}=\left(\rho,u,v,w,T\right)^T$ and perturbation quantities ${\boldsymbol{q}^\prime}=\left(\rho^\prime,u^\prime,v^\prime,w^\prime,T^\prime\right)^T$.
The base flow is assumed to be steady, spanwise invariant and local parallel, such that the perturbation growth is formulated by Linear Stability Theory (LST)\cite{mack1984boundarytransition}, based on the ansatz expressed as

\begin{equation}
\label{eq:LSTanstaz}
\boldsymbol{q}^{\prime}(x, y, z, t)=\hat{\boldsymbol{q}}(y) \exp [\mathrm{i}(\alpha x+\beta z-\omega t)]+\text { c.c., }
\end{equation}
where $\hat{\boldsymbol{q}}$ denote the 1-D perturbation shape functions in wall normal direction, $\alpha, \beta$ and $\omega$ are streamwise wavenumber, spanwise wavenumber and angular frequency, respectively. 
By inserting Eq.~(\ref{eq:LSTanstaz}) into the linearized Navier–Stokes (LNS) equations, the governing partial differential equations are reduced into a coupled system of second-order ordinary differential equations, with dependence only on the wall-normal coordinates:

\begin{equation}
\label{eq:CLSTsystem}
\boldsymbol{\mathcal{A}} \boldsymbol{\hat{q}}+\boldsymbol{\mathcal{B}}\frac{\mathrm{d}\boldsymbol{\hat{q}}}{\partial y}+\boldsymbol{\mathcal{C}}\frac{\mathrm{d}^2 \boldsymbol{\hat{q}}}{\partial y^2}=0
\end{equation}
where $\boldsymbol{\mathcal{A}}$,$\boldsymbol{\mathcal{B}}$, $\boldsymbol{\mathcal{C}}$ contains $\bar{\boldsymbol{q}}$, $\alpha,\beta$ and $\omega$.
A spatial stability formulation is employed in which the spanwise wavenumber $\beta$ and the angular frequency $\omega$ are specified a priori real number. The streamwise wavenumber $\alpha$ is unkown and complex, i.e. $\alpha=\alpha_r + i\alpha_i$, with its imaginary part quantifying the local rate of spatial amplification. Modal growth is therefore associated with negative values of $\alpha_i$.
Eq.~(\ref{eq:CLSTsystem}) is a generalized complex eigenvalue problem with quadratic terms in $\alpha$. Therefore, the companion matrix method\cite{danabasogluChebyshevMatrixMethod1990} is needed to formulate a linear problem with respect to $\alpha$, as expressed as

\begin{equation}
\label{eq:CLSTEVprob}
\boldsymbol{{A}}\boldsymbol{\tilde{q}}=\alpha \boldsymbol{{B}} \boldsymbol{\tilde{q}}
\end{equation}
where $\boldsymbol{\tilde{q}}=\left(\hat{\boldsymbol{q}},\alpha \hat{\boldsymbol{q}} \right)^T$ and
\begin{equation}
\boldsymbol{{A}}=
\left[
\begin{array}{cc}
\boldsymbol{\mathcal{A}} & -\boldsymbol{\mathcal{B}} \\
0 & \boldsymbol{I} 
\end{array}
\right],\quad
\boldsymbol{{B}}=
\left[
\begin{array}{cc}
0 & \boldsymbol{\mathcal{C}} \\
 \boldsymbol{I} & 0
\end{array}
\right]
\end{equation}
Once the local eigenvalue problems are all solved, the so-called $N$ factor is obtained by integrating the local growth rate along the streamwise direction:

\begin{equation}
N =\int_{x} - \alpha_i \mathrm{d} x
\end{equation}
When having multiple frequencies and spanwise wavenubmers, the envelope of $N$ factor is computed by

\begin{equation}
N_{\mathrm{env}}(x, \beta)=\max _\omega(N(x, \omega, \beta))
\end{equation}

\subsection{Transition Criteria With Wall-Temperature Effects} % Tomien Schlichting 
% Considering the FSC equations with boundary-layer temperature gradients and linear stability theory, we can obtain boundary-layer profiles and amplification factors for different temperature ratios ($T_w/T_e$), and pressure-gradient parameters ($\lambda_\theta$). Based on this extensive database, we can then derive correlations between the key transition-instability parameters (critical values) and their dominant influencing factors. For example, the original model correlated $Re_\mathrm{\theta t}$  only with $\lambda_\theta$ and Tu.
% Without loss of generality, we consider both wall heating and cooling scenarios.
% By additionally accounting for wall heating/cooling, we require a relation for $Re_\mathrm{\theta t}$ as a function of $\lambda_\theta$, Tu, and $T_w/T_e$.

% Revised version:
By combining the FSC equations, which account for boundary-layer temperature gradients, with linear stability theory, we calculate boundary-layer profiles and amplification factors across a range of temperature ratios ($T_w/T_e$) and pressure-gradient parameters ($\lambda_\theta$).
This extensive database enables the derivation of correlations between key transition-instability parameters (critical values) and their governing factors.
For instance, while the original model correlated $Re_\mathrm{\theta t}$ solely with $\lambda_\theta$ and Tu, the present framework incorporates wall temperature effects.
Without loss of generality, we consider both wall heating and cooling scenarios.
To account for temperature ratio effects, we need establish a relation for $Re_\mathrm{\theta t}$ as a function of $\lambda_\theta$, Tu, and $T_w/T_e$.

A critical component of the transition model is the local approximation of the momentum thickness Reynolds number $Re_{\theta}$.
In the original Langtry--Menter formulation, the ratio between the maximum vorticity Reynolds number ($\max(Re_{v,\max})$) and the momentum thickness Reynolds number ($Re_{\theta}$) acts as a key parameter ($(Re_{v,\max})/Re_{\theta} = 2.193$) but neglects the influence of pressure gradients. To improve accuracy across a wider range of aerodynamic configurations, \citet{franccois2023simplified} refined this correlation by explicitly accounting for pressure-gradient effects ($(Re_{v,\max})/Re_{\theta} =f (\lambda_\theta)$). Building on this approach, the present study further investigates and incorporates the influence of the wall-to-freestream temperature difference on this relationship ($(Re_{v,\max})/Re_{\theta} =f (\lambda_\theta, T_w/T_e)$).

% For crossflow-vortex-induced transition, the effect of temperature gradient must likewise be incorporated into the shape factor ($H_{12}^+$) and the crossflow-instability transition parameter ($Re_\text{He}^+$).

% \subsubsection{Correction for Tomien-Schlichting-Wave-Induced Transition}
% As described in Section~\ref{s:langtrymenter}, we need to correct the ration between the maximum strainrate Reynolds number ($Re_{v,\max}$) and the momentum thickness Reynolds number ($Re_{\theta}$), and the transition criteria $Re_\mathrm{\theta t}$ for the TS-wave-induced mechanisms to account for wall heating/cooling effects.
% The input parameters include the wall-temperature condition (expressed as the wall-to-ambient temperature ratio $T_w/T_e$), pressure gradient parameter ($\lambda_\theta$) at different Reynolds number (Up to $100\times10^6$).
% Based on the FSC equations and LST analysis, we obtain the envelope curves of TS-wave amplification at fixed $T_w/T_e \in [0.75, 1.25]$ and $m \in [-0.198, 1.0]$ over a range of Reynolds numbers.
% The $m$ and $\lambda_\theta$, which is utilized in the Langtry and Menter's model, has a relaption as follows:

\begin{figure}[h!]
	\centering
	\subfigure[The ratio of $Re_{v,\max}/Re_{\theta}$ vs $\lambda_\theta$ and $T_w/T_e$\label{f:MaxRev_Rett_Ratio_vs_LambdaTheta_Surface}]
	{\includegraphics[width=2.8in]{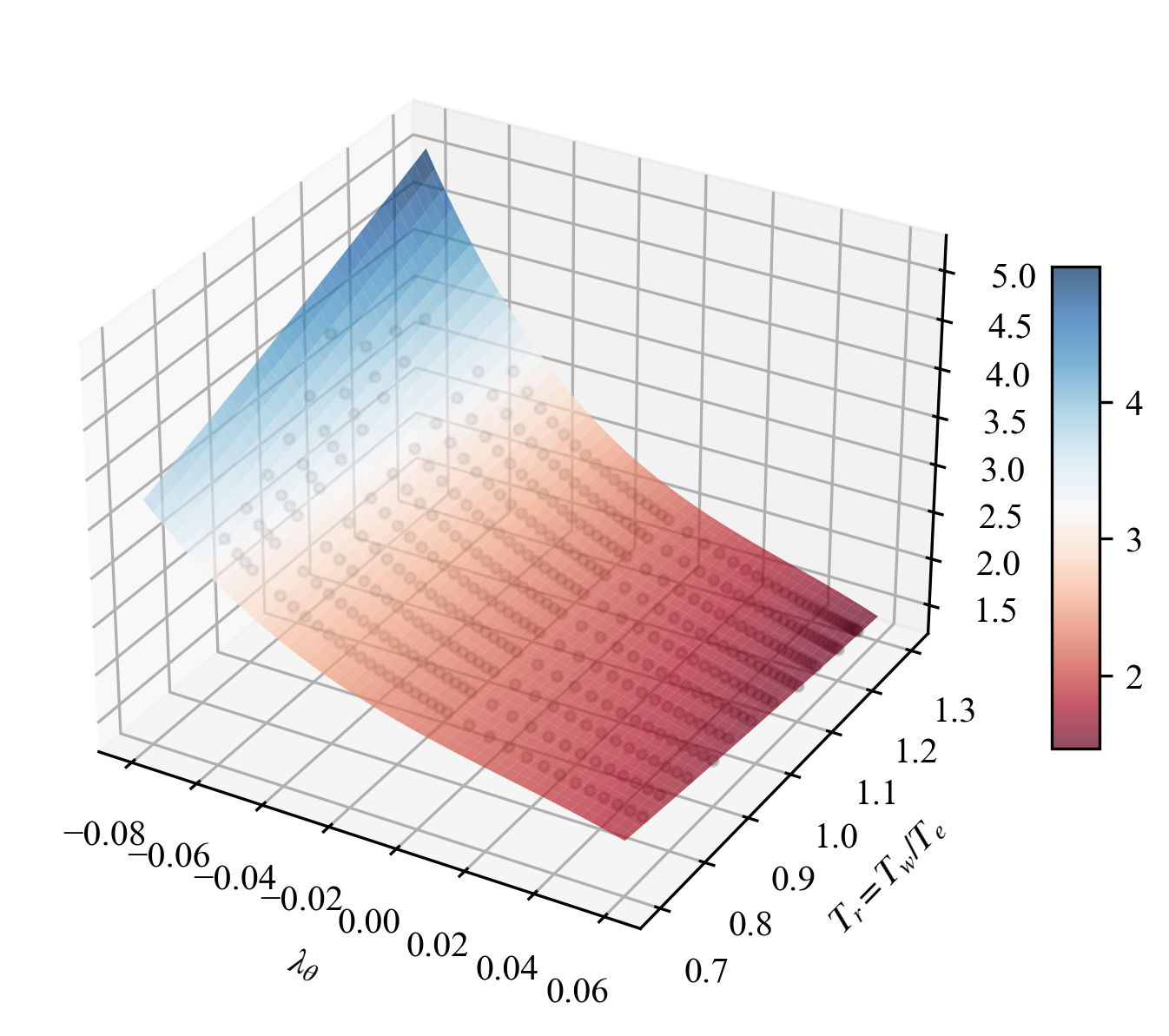}} 
	\subfigure[The ratio of $Re_{v,\max}/Re_{\theta}$ vs $\lambda_\theta$ at precribed temperature ratio \label{f:MaxRev_Rett_Ratio_vs_LambdaTheta_Tw}]
	{\includegraphics[width=2.8in]{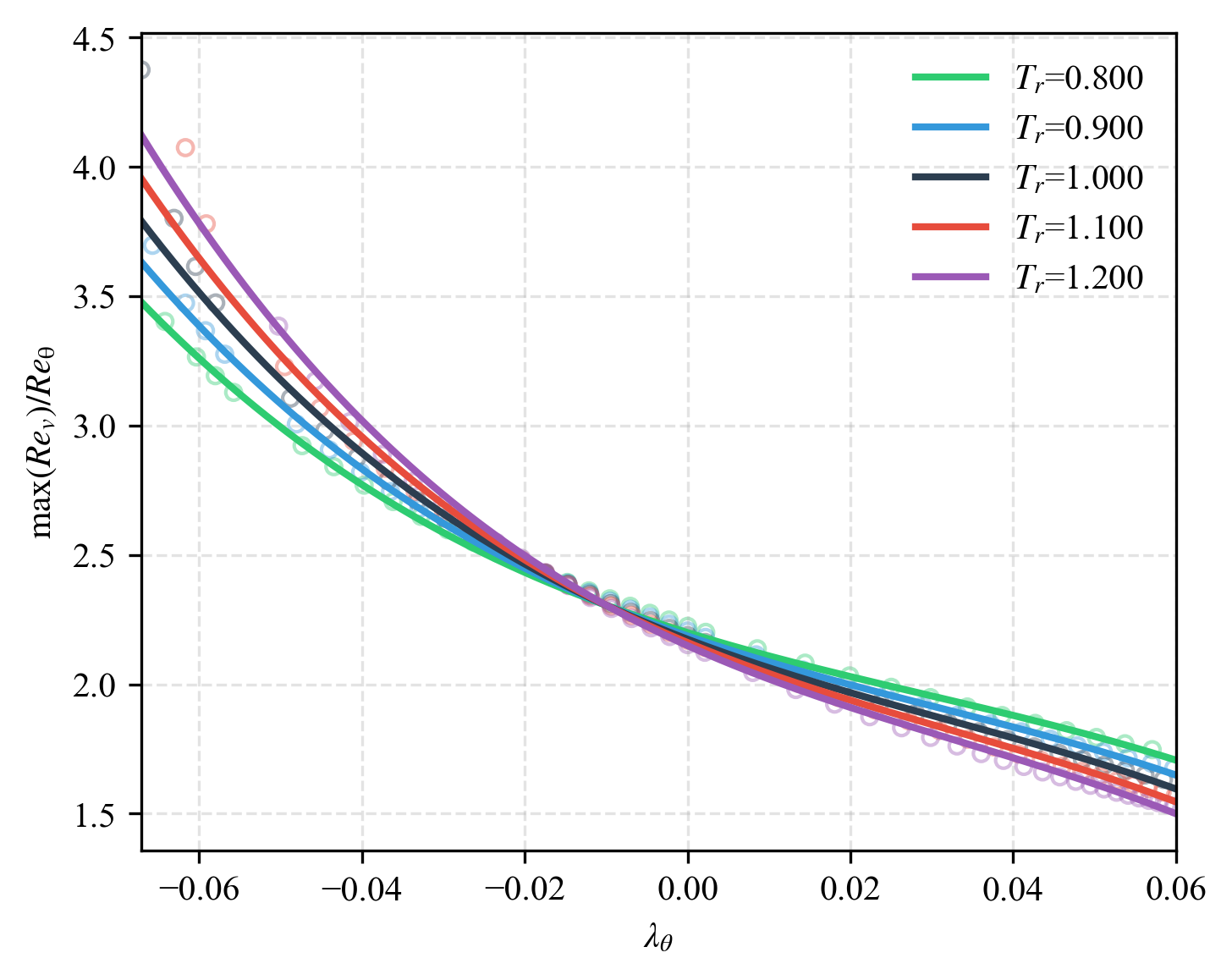}} 
	\caption{The description of $\bm{Re_{v,\max}/Re_{\theta}}$.}
	\label{f:MaxRev_Rett_Ratio}
\end{figure}

% Revised version:
In a summary, incorporating wall heating and cooling effects requires correcting two key model components: the ratio $\Pi^+$ between the maximum vorticity Reynolds number ($Re_{v,\max}$) and the momentum thickness Reynolds number ($Re_{\theta}$), and the transition criterion $Re_\mathrm{\theta t}$ for TS-wave-induced transition.
The FSC analysis considers the wall-temperature ratio ($T_w/T_e$) and the pressure-gradient parameter ($\lambda_\theta$) across a broad range of Reynolds numbers (up to $100\times10^6$).
Coupled with linear stability theory, we compute the envelope curves of TS-wave amplification over a broad range of operating conditions: temperature ratios $T_w/T_e \in [0.7, 1.3]$, and parameters $m \in [-0.198, 1.0]$.
The parameter $m$ is related to the pressure-gradient parameter $\lambda_\theta$ employed in the transport model by the following equation:
\begin{equation}	
	\lambda_\theta=\frac{\theta^2 m  u_{e}}{ \nu_e x}.
\end{equation}
The ratio $Re_{v,\max}/Re_{\theta}$ can be determined from boundary layer information. Fig.~\ref{f:MaxRev_Rett_Ratio} (a) presents the contour plot of $(Re_{v,\max})/Re_{\theta}$ with respect to $T_w/T_e$ and $\lambda_\theta$, while Fig.~\ref{f:MaxRev_Rett_Ratio} (b) shows the variation of $(Re_{v,\max})/Re_{\theta}$ with $\lambda_\theta$ at different $T_w/T_e$ ($T_r$).
In Fig.~\ref{f:MaxRev_Rett_Ratio}, the circular scatter points represent the data obtained based on the FSC equations, while the surface or solid lines are the fitted results.
It can be observed that under heating conditions ($T_r > 1$), the ratio $\Pi^+$ increases with rising temperature ratio, whereas under cooling conditions, $\Pi^+$ decreases as the temperature ratio drops.
We correct the ratio $\max(Re_{v,\max})/Re_{\theta}$ as follows:
\begin{equation}
	\label{eq:MaxRev_Rett_Ratio}	
	\begin{aligned}
	\frac{Re_{v,\max}}{Re_{\theta}} = \Pi^+ = {}&
	\left(2.312813 + 0.8074101\lambda_\theta + 6.572101\lambda_\theta^2 - 352.2001\lambda_\theta^3 + 3.143236\lambda_\theta^4\right) \\
	&+ T_r\left(-0.1514422 - 14.56479\lambda_\theta + 54.26127\lambda_\theta^2 - 368.1236\lambda_\theta^3 + 3.342443\lambda_\theta^4\right) \\
	&+ T_r^2\left(0.01279790 + 1.768478\lambda_\theta + 44.95197\lambda_\theta^2 - 391.6507\lambda_\theta^3 + 3.394677\lambda_\theta^4\right).
	\end{aligned}
\end{equation}

\begin{figure}[h!]
	\centering
	\subfigure[$\beta=0$ ($\lambda_\theta =0$), $\text{Tu} = 0.09\%$ ($N_\mathrm{crit} = 8.40$) \label{f:LambdaThetaN0p0}]
	{\includegraphics[width=2.8in]{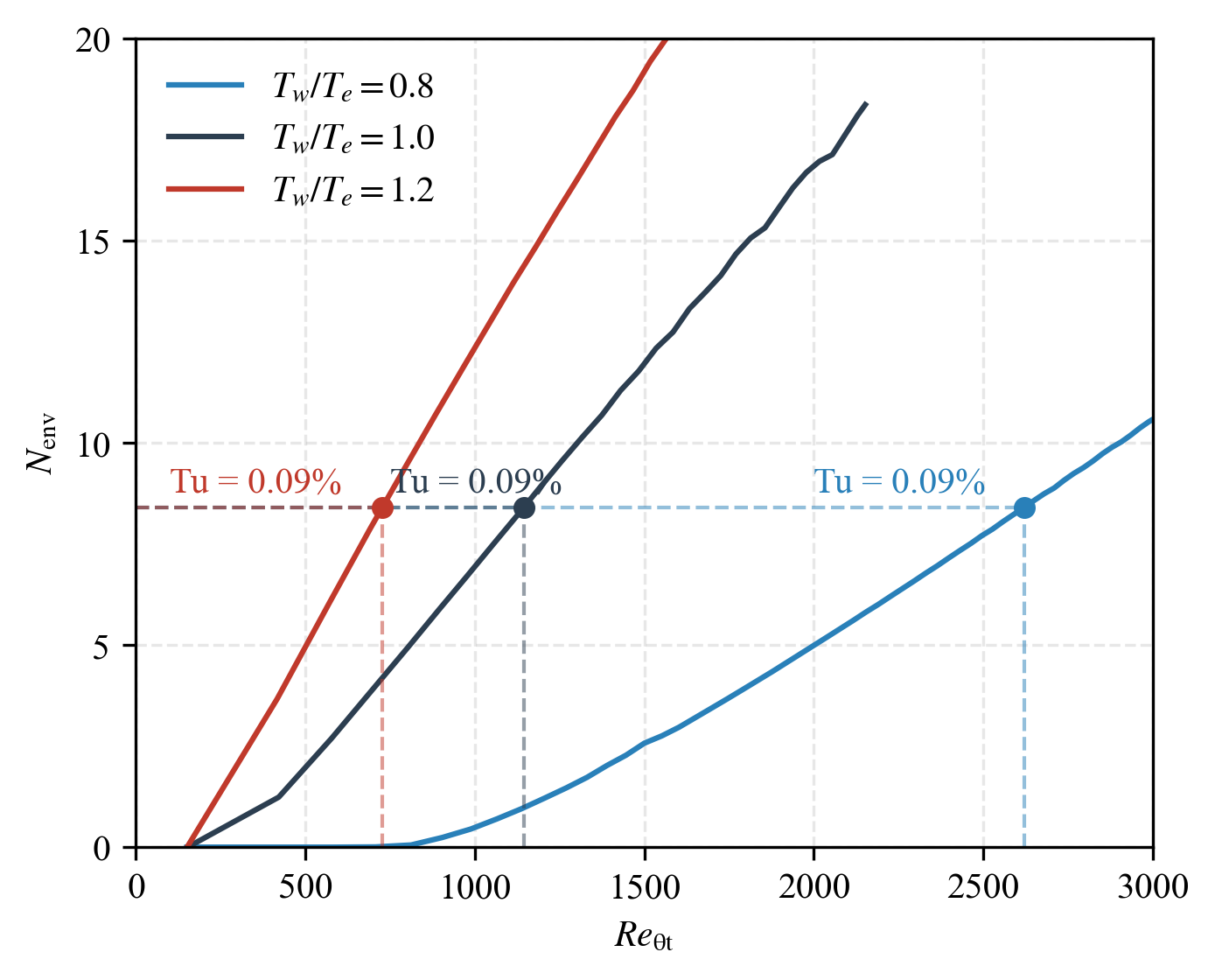}} 
	\subfigure[$\beta=-0.13$($\lambda_\theta =-0.0368$), $\text{Tu} = 0.03\%$ ($N_\mathrm{crit} = 11.04$)\label{f:LambdaThetaN0p5}]
	{\includegraphics[width=2.8in]{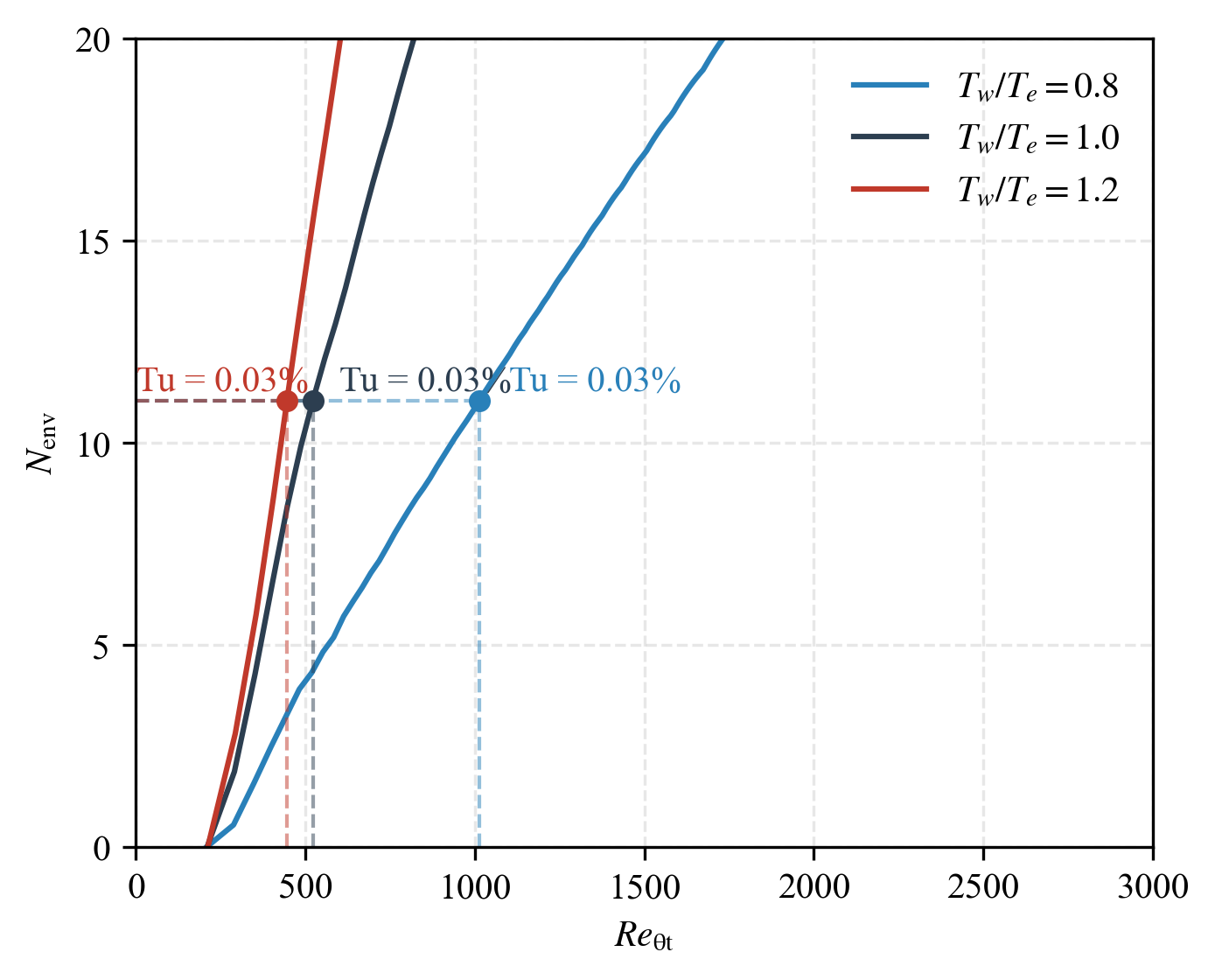}} 
	\caption{The envelope curves of TS-wave amplification at fixed $\bm{T_w/T_e}$ and $\bm{\beta}$ over a range of momentum Reynolds number.}
	\label{f:TSNRett}
\end{figure}
% Tu, Mack关系，说一下ratio比率

Based on different Tu, we can determine the critical momentum Reynolds number $Re_\mathrm{\theta t}$ at transition location by using LST analysis and the $e^N$ method (Fig.~\ref{f:TSNRett}).%Figure.~\ref{}
The $N_\text{crit}$ is determined by the turbulence intensity (Tu) by using Mack's relationship~\cite{mack1984boundarytransition}
\begin{equation}
\label{eq:Ncrit_Tu}
N_\mathrm{crit} = -8.43 -2.4 \ln(\mathrm{Tu}) .
\end{equation}
In Fig.~\ref{f:TSNRett}, the envelope curves of TS-wave amplification at fixed $T_w/T_e$ and $\lambda_\theta$ over a range of momentum Reynolds number are presented.
We can see that with the increase of wall temperature ratio ($T_w/T_e$), the transition location moves upstream, indicating that wall heating has a destabilizing effect on the boundary layer.
The wall cooling has the opposite effect, and the critical $Re_\mathrm{\theta t}$ increase significantly.
Meanwhile, with the increase of adverse pressure gradient (more negative $\lambda_\theta$), the transition location also moves upstream, indicating that the adverse pressure gradient has a destabilizing effect on the boundary layer.

\begin{figure}[h!]
	\centering
	\subfigure[Tu = 0.05\%\label{f:Retc_vs_lambda_theta_fit_Tu0.001}]
	{\includegraphics[width=2.8in]{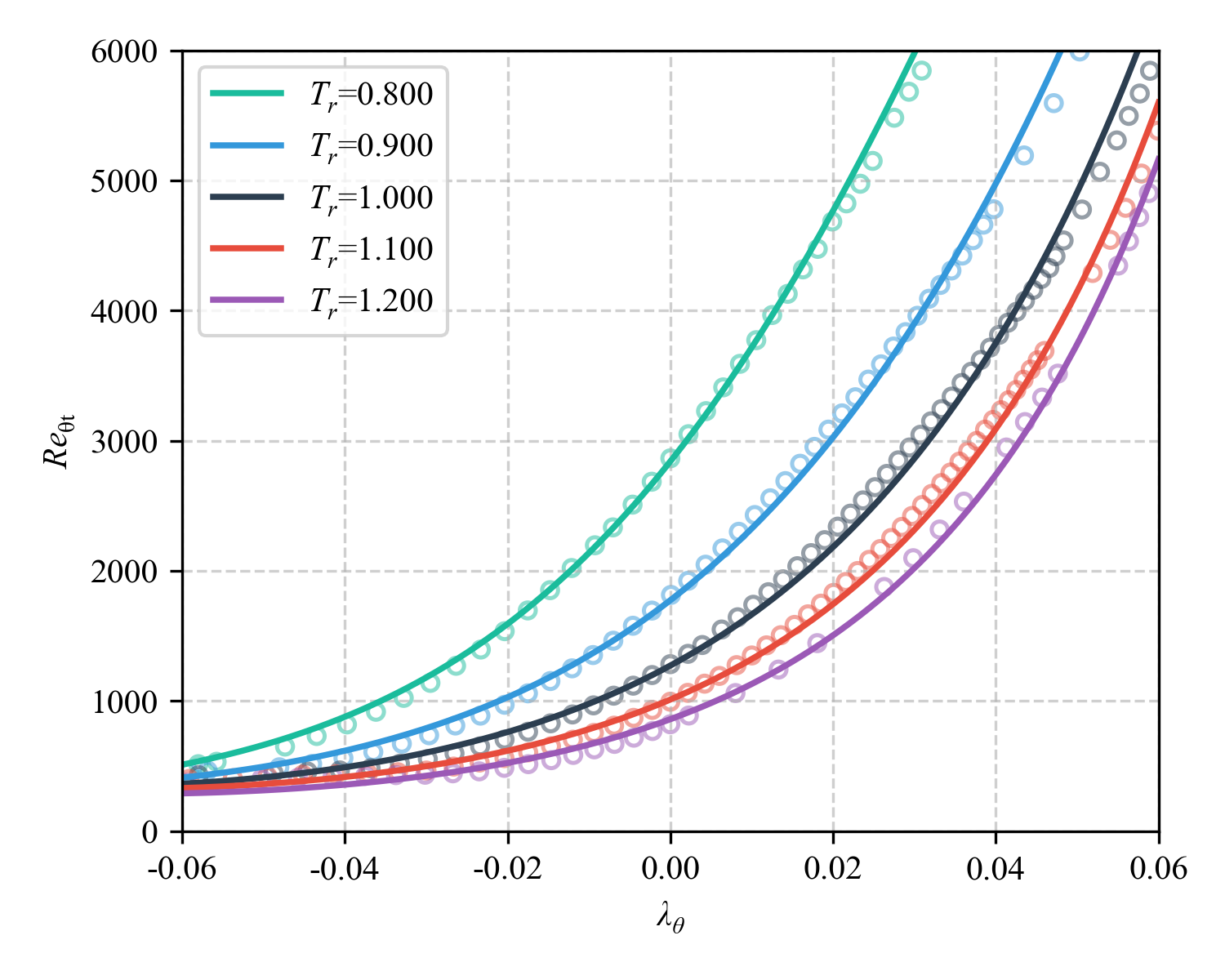}} 
	\subfigure[Tu = 0.1\%\label{f:Retc_vs_lambda_theta_fit_Tu0.0005}]
	{\includegraphics[width=2.8in]{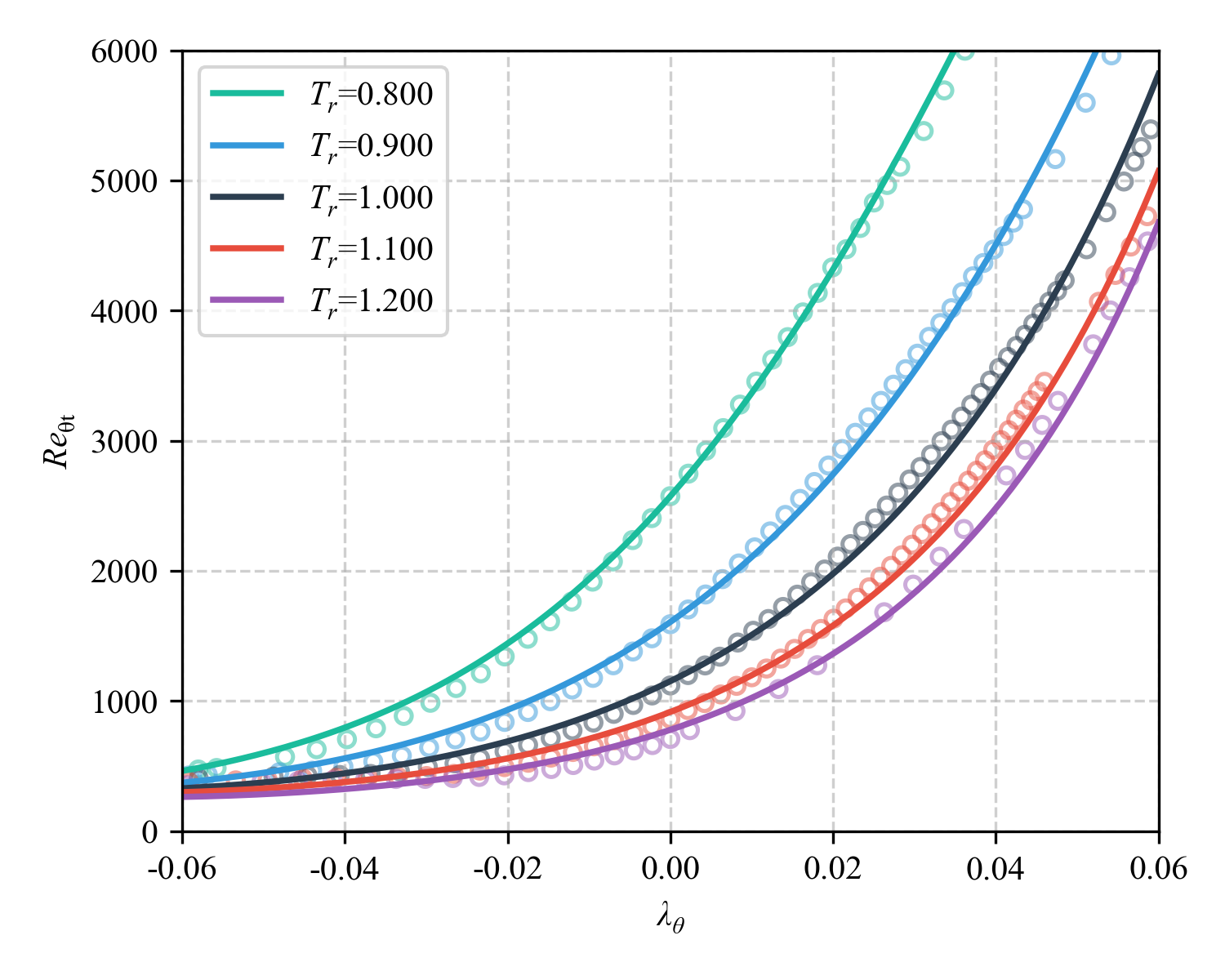}} 
	\caption{Variation of $\bm{Re_{\theta t}}$ with $\bm{\lambda_\theta}$ for prescribed $\bm{T_r}$ at different turbulence intensities.}
	\label{f:Retc_vs_lambda_theta}
\end{figure}

\begin{figure}[h!]
	\centering
	\subfigure[Tu = 0.05\%\label{f:Retc_surafce_Tu0.000400}]
	{\includegraphics[width=2.8in]{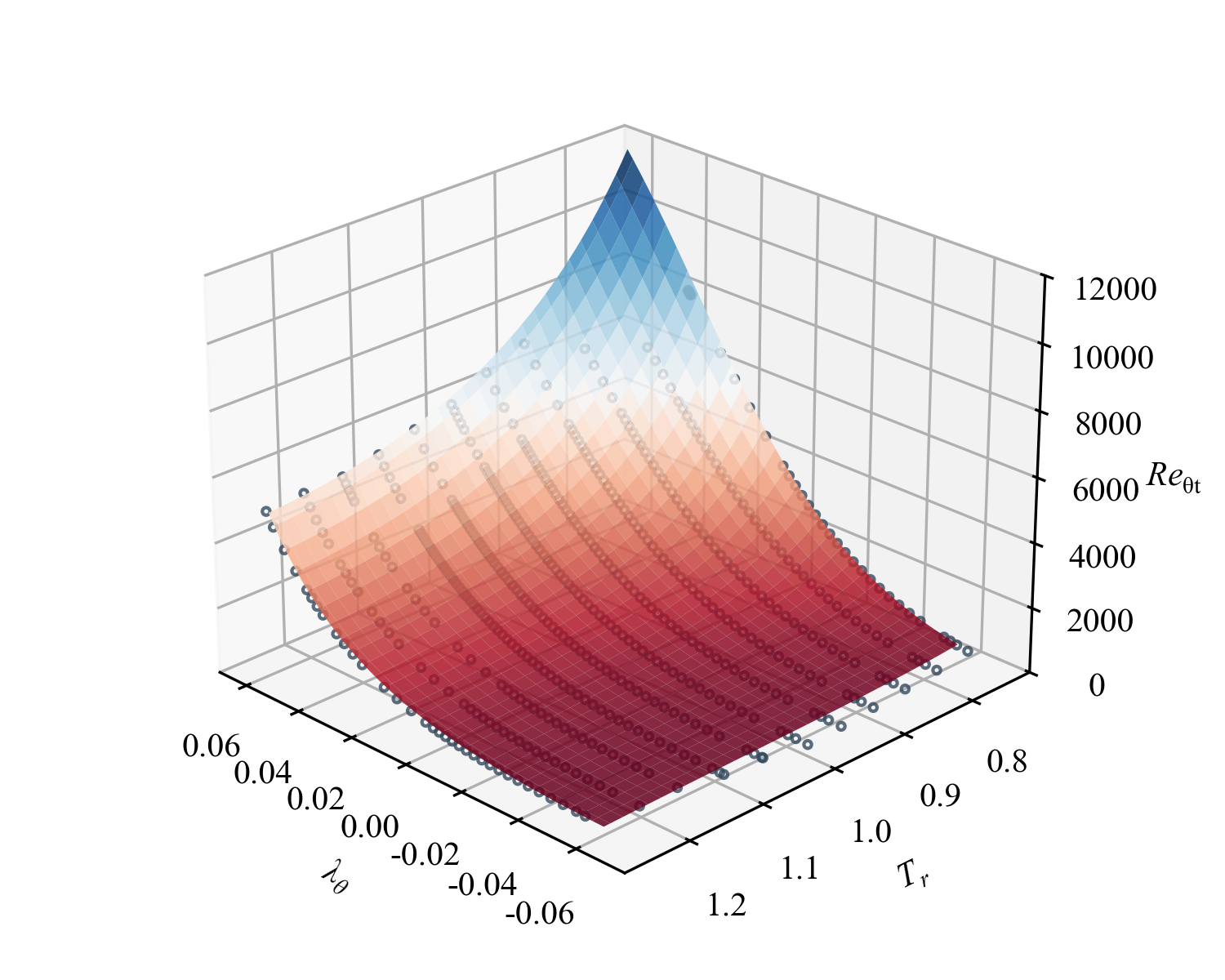}} 
	\subfigure[Tu = 0.1\%\label{f:Retc_vs_lambda_theta_fit_Tu0.000800}]
	{\includegraphics[width=2.8in]{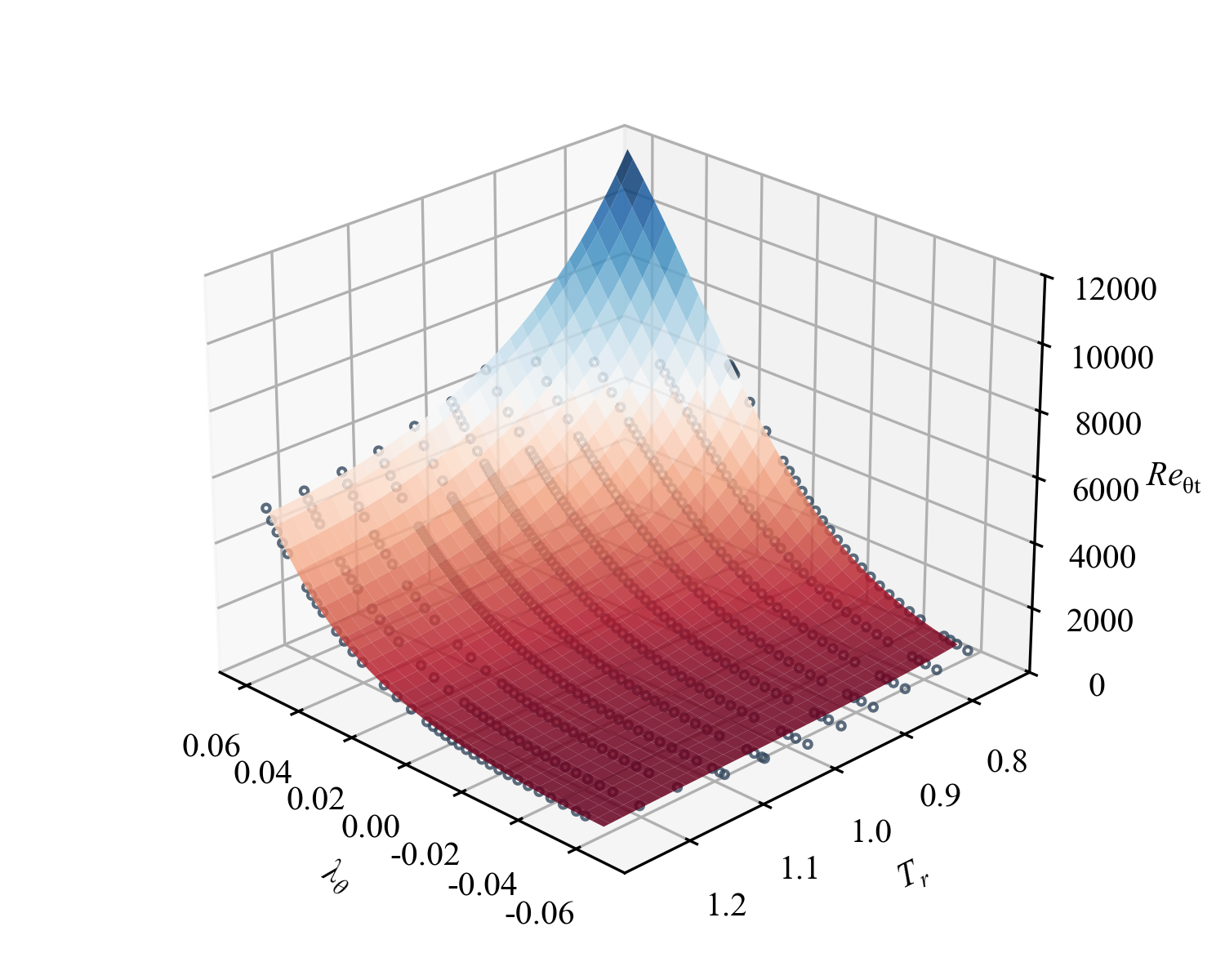}} 
	\caption{The surface of $\bm{Re_{\theta t}}$ with $\bm{\lambda_\theta}$ and $\bm{T_r}$ at different turbulence intensities.}
	\label{f:Retc_surface}
\end{figure}

We have $4024\times600$ (4024 corresponds to the combinations of temperature ratios, Tu $\in [0.0003,0.003]$ and $\beta$ ($\lambda_\theta$) values, while the 600 refers solely to the distinct local Reynolds numbers) data points to fit the correction function of $Re_\mathrm{\theta t}$ (the momentum thickness based Reynolds number at the transition location) with Tu, $T_w/T_e$ and $\lambda_\theta$.
The results are shown in Fig.~\ref{f:Retc_vs_lambda_theta} and \ref{f:Retc_surface}.
Fig.~\ref{f:Retc_vs_lambda_theta} shows the variation of $Re_{\theta t}$ with $\lambda_\theta$ for prescribed $T_r$ at different turbulence intensities, while Fig.~\ref{f:Retc_surface} presents the surface of $Re_{\theta t}$ with $\lambda_\theta$ and $T_r$ at different turbulence intensities.
We show the results for two turbulence intensities (Tu = 0.05\% and Tu = 0.1\%) as examples.
The circular scatter points represent the data obtained based on the FSC equations and LST stability analysis, while the curves and surfaces are the fitted results.
The fitting accuracy is high, with $\mathcal{R}^2$ values exceeding 0.998 in all cases.
We can see that the effect of cooling is significantly greater than that of heating. The influence of heating tends to saturate, regardless of whether the pressure gradient is adverse ($\lambda_\theta < 0$) or favorable ($\lambda_\theta > 0$).
In general, $Re_{\theta t}$ increases as turbulence intensity decreases.
It should be noted that the sensitivity of transition to wall temperature is considerably reduced under adverse pressure gradient conditions.

Finally, we obtain the correction function of $Re_\mathrm{\theta t}$ as follows:
\begin{equation}
\label{eq:ReThetaT_Correction}
	\ln (Re_\mathrm{\theta t}) = \ln (h(\lambda_\theta, T_r)) + \ln (h(\mathrm{Tu})), 
\end{equation}
where $\ln (h(\lambda_\theta, T_r)) $ is expressed as:
\begin{equation}
\label{eq:ReThetaT1}
	\begin{aligned}
	\ln (h(\lambda_\theta, T_r)) =\;&
	 \left(20.905432 - 34.936088\,\lambda_\theta - 1162.249196\,\lambda_\theta^2 - 792.760934\,\lambda_\theta^3 + 5319.967633\,\lambda_\theta^4\right)  \\
	&+ T_r\left(-30.666453 + 210.612461\,\lambda_\theta + 1992.846133\,\lambda_\theta^2 + 45.368047\,\lambda_\theta^3\right) \\
	&+ T_r^2\left(22.766542 - 230.627111\,\lambda_\theta - 798.668631\,\lambda_\theta^2\right) \\
	& + T_r^3\left(-5.872553 + 81.554132\,\lambda_\theta\right) ,
	\end{aligned}
\end{equation}
and $\ln (h\left(\mathrm{Tu}\right))$ is defined as:
\begin{equation}
	\label{eq:TuCorrectionFactor}
	\ln\!\left(h\left(\mathrm{Tu}\right)\right) = 153.768618
	+ 82.899363\,\ln\!\left(\mathrm{Tu}\right)
	+ 16.673825\,\left[\ln\!\left(\mathrm{Tu}\right)\right]^2
	+ 1.485418\,\left[\ln\!\left(\mathrm{Tu}\right)\right]^3
	+ 0.049537\,\left[\ln\!\left(\mathrm{Tu}\right)\right]^4 .
\end{equation}
It can be observed that $h\left(\mathrm{Tu}\right)$ is a monotonically decreasing function.

% 画出一些图，比如不同温度比下，Rev/Tett的变化曲线？
% 不同温度下，某些边界层特性和LambdaTheta的关系曲线？说明不受绝热温度的影响？？
% 也要画出不同压力梯度下，指定Tu，Rett和LambdaTheta的关系曲线,说明变化趋势。
% \subsubsection{Correction for Crossflow-Vortex-Induced Transition} % Tomien Schlichting 
% For the CF-vortex-induced transition, we need to correct the shape factor ($H_{12}^+$) and the crossflow-instability transition parameter ($Re_\text{He}^+$) to account for wall heating/cooling effects.
% The input parameters include $T_w/T_e$, $\lambda_\theta$, at different Reynolds number (Up to $12\times10^6$) and sweep angle ($\Lambda$).
% Based on the FSC equations and LST analysis, we obtain the envelope curves of CF-vortex amplification at fixed $T_w/T_e$, $\lambda_\theta$ and $\Lambda$ over a range of Reynolds numbers.

% \section{Improved Stability-based Transition Modeling Considering Wall-to-freestream temperature effects and Validation}

\section{Transport Transition Modeling and Validation}
\label{s:transportequation}
\subsection{The Improved Stability-based Transition Transport Model} % Tomien Schlichting
We used the stability-based transition modeling originally proposed by \citet{franccois2023simplified} as the base model.
The transport equation of the intermittency $\gamma$ in this modified model is expressed as:
\begin{equation}
	\frac{\partial(\rho \gamma)}{\partial t}+\frac{\partial\left(\rho u_j \gamma\right)}{\partial x_j}=P_\gamma-E_\gamma+\frac{\partial}{\partial x_j}\left[\left(\mu+\frac{\mu_t}{\sigma_f}\right) \frac{\partial \gamma}{\partial x_j}\right],
\end{equation}
where $P_\gamma$ and $E_\gamma$ are the production and destruction/relaminarization terms, respectively.
$\sigma_f$ is set to 1.0.
The $\mu$ and$\mu_t$ are the laminar and turbulent dynamic viscosities, respectively.
% To keep the paper compact and concise, only the key parameters and variables are listed here; definitions of the remaining quantities can be found in \citet{Langtry2009}.
%  and \cite{Grabe2018}
The production term is given as:
% \begin{equation}
% 	P_{\gamma, \mathrm{ts}, \mathrm{cw}}=\left(F_{\text {length }} \right) \rho {S} \gamma \left(1-\gamma\right) (F_\mathrm{onset, ts} + F_\mathrm{onset, cf}),
% \end{equation}
\begin{equation}
	P_{\gamma}=F_{\text{length}}  \rho {S} \left(1- \gamma\right),
\end{equation}
% where $F_\text{length}$ is a constant value of 100.0.
where $F_{\text{length}}$ indicates the production rate of $\gamma$ and is set as 14.0.
% The destruction source is idential to the one used in the original Langtry-Menter model.
% $C_\text{a1}$ and $C_\text{e2}$ are 2.0 and 50.0, respectively.
${S}$ is the magnitude of the strain rate tensor.
$F_\mathrm{onset}$ activates the production term when the local flow reaches the transition onset condition.
In this study, we improve the definition of $F_\mathrm{onset}$ to consider wall-temperature effects, and it is defined as:
% \begin{equation}
% 	\begin{aligned}
% 		&F_{\mathrm{onset}}=\max \left(F_{\mathrm{onset} 2}-F_{\mathrm{onset} 3}, 0.0\right), \quad F_{\mathrm{onset3}}=\max \left(1-\left(\frac{R_T}{2.0}\right)^3, 0.0\right) ; \quad R_T=\frac{\mu_t}{\mu} \\
% 		&F_{\text {onset2 }}=\min \left(\max \left(F_{\text {onset1 }}, F_{\text {onset1 }}^4\right), 2.0\right), F_{\text {onset1 }}=\frac{R e_\theta^*}{R e_{\theta t}} ; \quad R_\theta^*=\frac{R e_\nu}{\pi} ; \quad R e_\nu=\frac{d^2}{\nu} S, 
% 	\end{aligned}
% \end{equation}
% and $\pi$ is defined as:
% \begin{equation}
% 	\pi=0.071665 H_{12}^3-0.73186 H_{12}^2+4.2563 H_{12}-5.1743.
% \end{equation}
\begin{equation}
	\label{eq:fonset}
	\begin{aligned}
		& F_{\text {onset}1}=\frac{\operatorname{Re}_{v}}{\Pi^+ \operatorname{Re}_{\theta t}}, \quad F_{\text {onset}2}=\min \left(\max\left(F_{\text {onset} 1}, F_{\text {onset} 1}^4\right), 2.0\right)  \\
		& F_{\text {onset}3}=\max \left(1-\left(\frac{R_T}{2.0}\right)^3, 0.0\right), \quad F_{\text {onset}}=\max \left(F_{\text {onset}2}-F_{\text{onset}3}, 0.0\right) \\
		& R_T=\frac{\mu_t}{\mu }, \quad \operatorname{Re}_{{v}}=\frac{\rho d^2 S}{\mu}
	\end{aligned},
\end{equation}
where $F_{\mathrm{onset}1}$ determines whether the local flow has reached the transition onset condition.
Here, $Re_{\mathrm{\theta t}}$ is calculated using Eqs.~\eqref{eq:ReThetaT_Correction}, \eqref{eq:ReThetaT1}, and \eqref{eq:TuCorrectionFactor} in Section~\ref{s:modelformulation}.
It depends on the freestream turbulence intensity (Tu), the pressure gradient parameter ($\lambda_\theta$), and the wall-to-freestream temperature ratio ($T_w/T_e$).
The term $Re_{v}/\Pi^+$ represents the local momentum thickness Reynolds number $Re_\theta$, as defined in Eq.~\eqref{eq:MaxRev_Rett_Ratio}.
In Eq.~\eqref{eq:fonset}, $Re_v$ is the vorticity Reynolds number, $d$ is the distance to the nearest wall, and the wall-to-freestream temperature ratio ($T_w/T_e$) is considered.

The pressure gradient parameter $\lambda_\theta$ is defined as:
\begin{equation}
	\label{eq:lambdatheta}
	\lambda_\theta=\frac{\rho \theta^2}{\mu} \frac{d Q_e}{d s} = {Re}_\mathrm{\theta}^2 \frac{\nu_e}{Q_e^2} \frac{d Q_e}{d s} ,
\end{equation}
where $ {Re}_\mathrm{\theta}$ is computed by $ (Re_{v}) / \Pi^+$~\eqref{eq:MaxRev_Rett_Ratio}.
In Eq.~\eqref{eq:lambdatheta}, $\dif Q_e /\dif s$ is given by
\begin{equation}
	\frac{d Q_e}{d s}= \frac{u}{Q_e} \frac{\dif Q_e}{\dif x} + \frac{v}{Q_e} \frac{\dif Q_e}{\dif y} + \frac{w}{Q_e} \frac{\dif Q_e}{\dif z} ,
\end{equation}
where $Q_e$ is the magnitude of the velocity vector at the boundary-layer edge and computed from the compressible Bernoulli equation and is expressed as:
\begin{equation}
	Q_e=\sqrt{U_{\infty}^2+\frac{2 \kappa_r}{\kappa_r-1}\left[1-\left(\frac{p}{p_{\infty}}\right)^{(\kappa_r-1) / \kappa_r}\right] \frac{p_{\infty}}{\rho_{\infty}}},
\end{equation}
where $\kappa_r$ is the ratio of specific heats.
% Note that we do not consider the compressible effects ($M_e$) in computing $Re_{\theta t}$.
The edge-velocity gradients in the three coordinate directions are calculated using the external potential flow pressure gradients as follows:
\begin{equation}
	\begin{aligned}
		& \frac{\dif Q_e}{\dif x} = -\frac{1}{\rho_\infty Q_e} \frac{p}{p_\infty}^{-1/\kappa_r} \frac{\dif p}{\dif x} ,\\
		& \frac{\dif Q_e}{\dif y} = -\frac{1}{\rho_\infty Q_e} \frac{p}{p_\infty}^{-1/\kappa_r} \frac{\dif p}{\dif y} ,\\
		& \frac{\dif Q_e}{\dif z} = -\frac{1}{\rho_\infty Q_e} \frac{p}{p_\infty}^{-1/\kappa_r} \frac{\dif p}{\dif z}  .
	\end{aligned}		
\end{equation}
As the formulation of the edge-velocity gradient in the streamwise direction is not Galiean-invariant, it can be only applied to fixed-wall configurations.
% We found that the temperature ratio has limited effects on critical ratio values of ($Re_v/2.193/Re_\mathrm{\theta t}$).

We improve $\Pi^+$ as Eq.~\eqref{eq:MaxRev_Rett_Ratio} to consider the temperature difference.
We can see the temperature difference has a significant effect on $\Pi^+$ (Fig.~\ref{f:MaxRev_Rett_Ratio}), which further influences the transition onset function $F_\text{onset}$.
The $T_w$ and $T_e$ are calculated by the local temperature and freestream temperature (the unit is $K$).
In Eq.~\eqref{eq:fonset}, $F_\text{onset2}$ ensures a rapid growth of intermittency once the transition onset condition is met, while $F_\text{onset3}$ accounts for the damping effect of existing turbulence in the boundary layer.
The destruction/relaminarization term $E_\gamma$ is expressed as:
\begin{equation}
	E_\gamma= c_{a 2} \rho \Omega \gamma F_{\text {turb }}\left(c_{e 2} \gamma-1\right),
\end{equation}
where $C_\text{a2}$ and $C_\text{e2}$ are model constants set to 1.0 and 50.0, respectively.
The $\Omega$ is the magnitude of vorticity, and $F_\text{turb}$ is a function that activates the destruction term in turbulent regions, given by
\begin{equation}
	F_{\mathrm{turb}}=\exp \left[-\left(\frac{R_T}{4}\right)^4\right].
\end{equation}

The intermittency $\gamma$ is then used to modify the production term of the $k-\omega$ shear-stress transport (SST) turbulence model~\cite{Menter1994,menter2003ten}, effectively activating the turbulence production in the transition and turbulent regions.
It consists in modifying the production term of the turbulent kinetic energy ($k$) equation as follows:
\begin{equation}	
	\frac{\partial}{\partial}(\rho k)+\frac{\partial}{\partial x_j}\left(\rho u_j k\right)=\widetilde{P}_k-\widetilde{D}_k+\frac{\partial}{\partial x_j}\left[\left(\mu+\sigma_k \mu_t\right) \frac{\partial k}{\partial x_j}\right],
\end{equation}
where $\widetilde{P}_k=\gamma P_k$ and $\widetilde{D}_k=\min [\max (\gamma, 0.1), 1.0] D_k$.
The production of $k$ is suppressed in laminar and transitional regions by the intermittency $\gamma,$ while the destruction term is only partially suppressed in laminar regions to avoid numerical issues.
When $\gamma$ is 1.0, the original $k-\omega$ SST model is recovered.

\subsection{Comparison of the Improved Model with LST-based Approch for Typical Cases}
\begin{figure}[h!]
  \centering
  \includegraphics[width=0.7\textwidth]{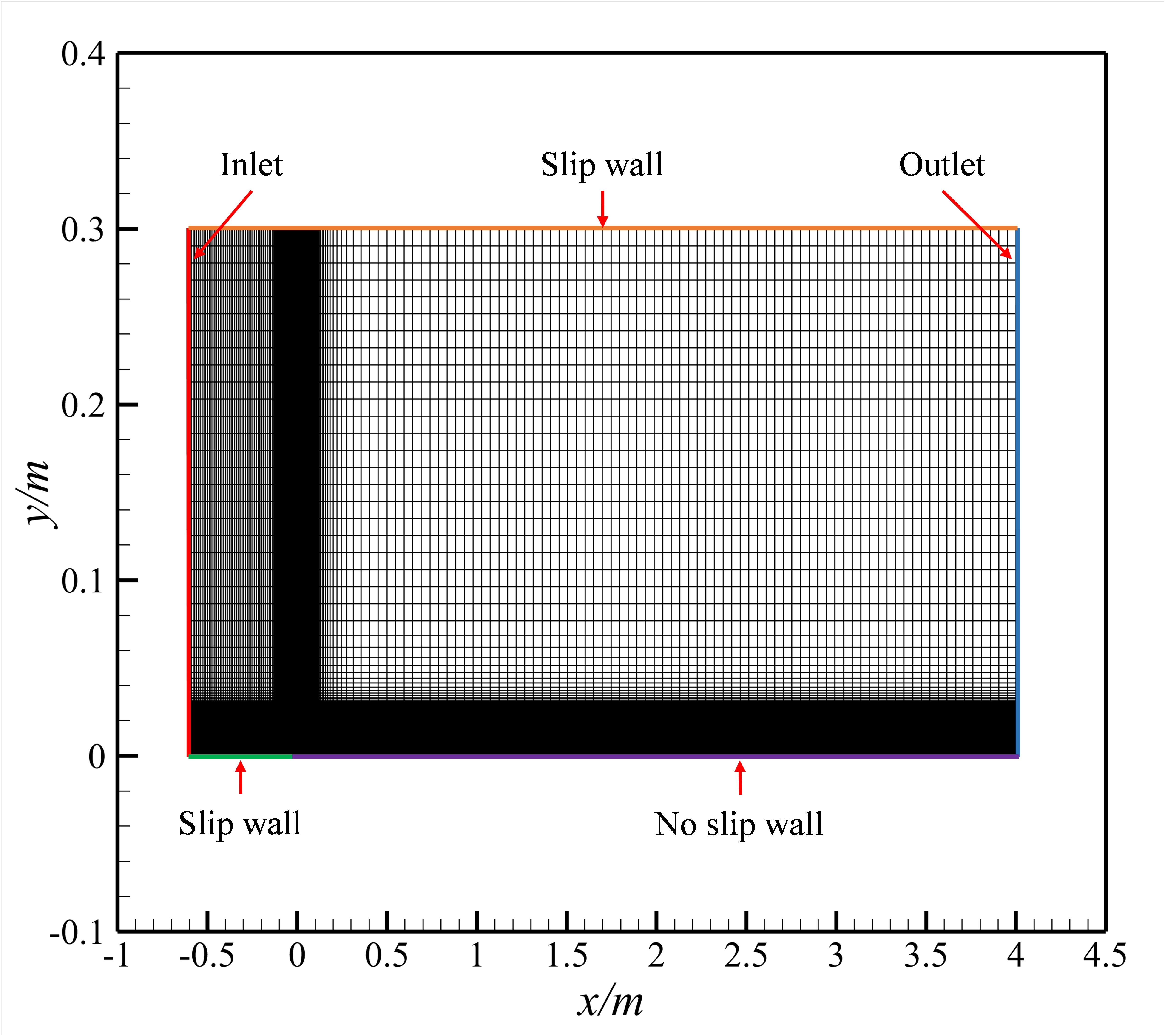}
  \caption{The flat plate computational grid for the SK case.}
  \label{f:SKMesh}
\end{figure}

% \subsubsection{Schubauer and Klebanof Flate Plate} %就以平板为例
% 我们选择用SK平板，来对比标定模型和LST的结果。SK平板是经典的由TS波诱导的转捩实验。其速度、湍流度、单位雷诺数是。本文用于计算的网格见，Y+是。我们给出了不同温度比下，标定模型和LST的对比结果。可以看出，标定模型能够很好地捕捉温度对转捩位置的影响趋势。图中的圆圈，代表的绝热壁温度下的实验结果。
We chose the Schubauer and Klebanoff (SK) flat plate to compare the results of the improved model with those from LST~\cite{schubauer1956contributions}. 
The SK flat plate is a classic experiment of transition induced by TS instabilities.
The velocity is 50$m/s$, the turbulence intensity is 0.1\%, and unit Reynolds number is $3.34\times10^6$.
The computational grid used in this study and the boundary conditions are are shown and set as Fig.~\ref{f:SKMesh}.
The grid mesh consists of 361 points in the streamwise direction and 161 points in the normal direction, with $y^+$ values of 0.2 and growth ratio of 1.12.

\begin{figure}[h!]
  \centering
  \includegraphics[width=0.5\textwidth]{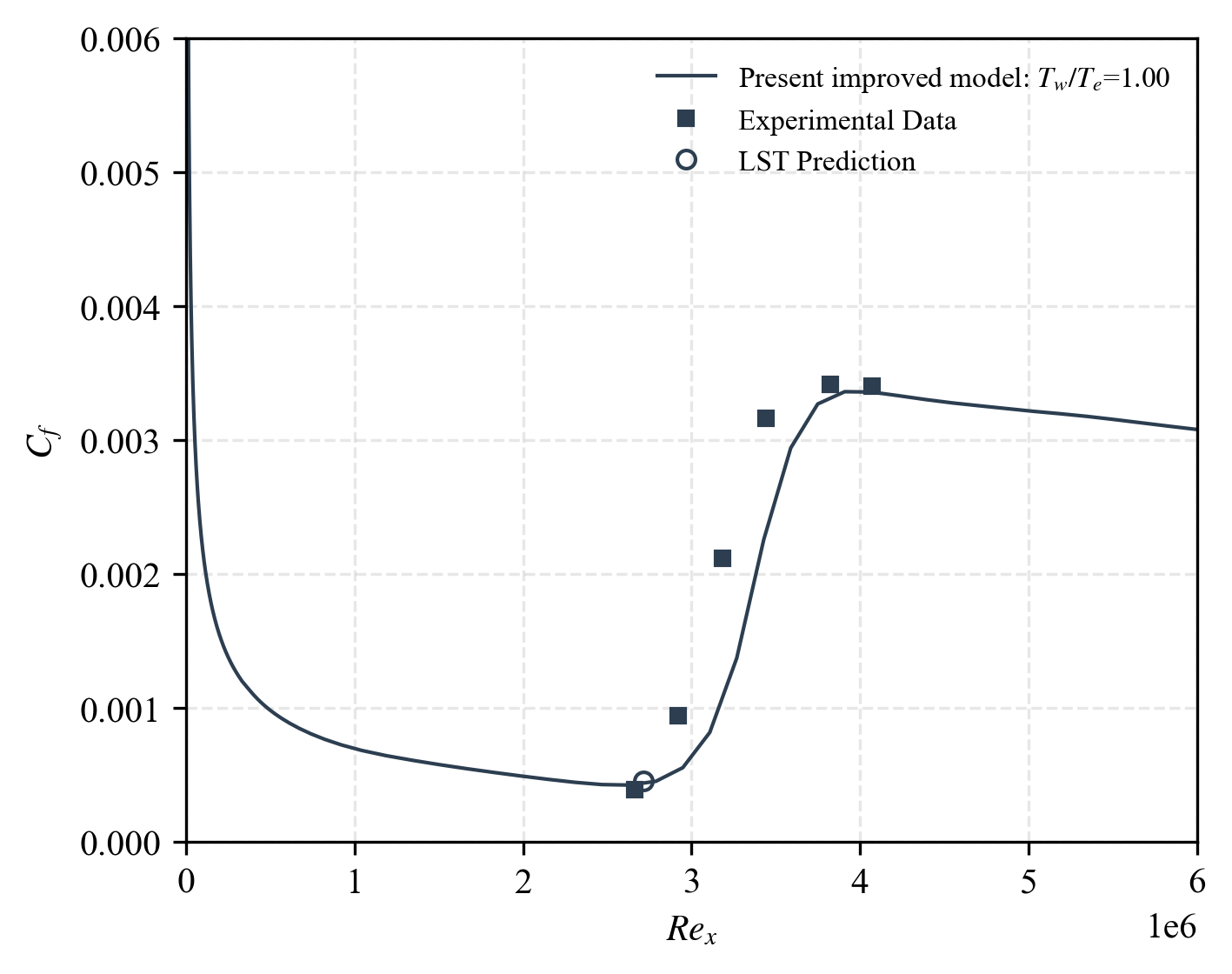}
  \caption{Comparison of skin friction among the improved model, LST, and experimental results at the adiabatic wall temperature.}
  \label{f:SKCompare}
\end{figure}
\begin{figure}[h!]
  \centering
  \includegraphics[width=0.5\textwidth]{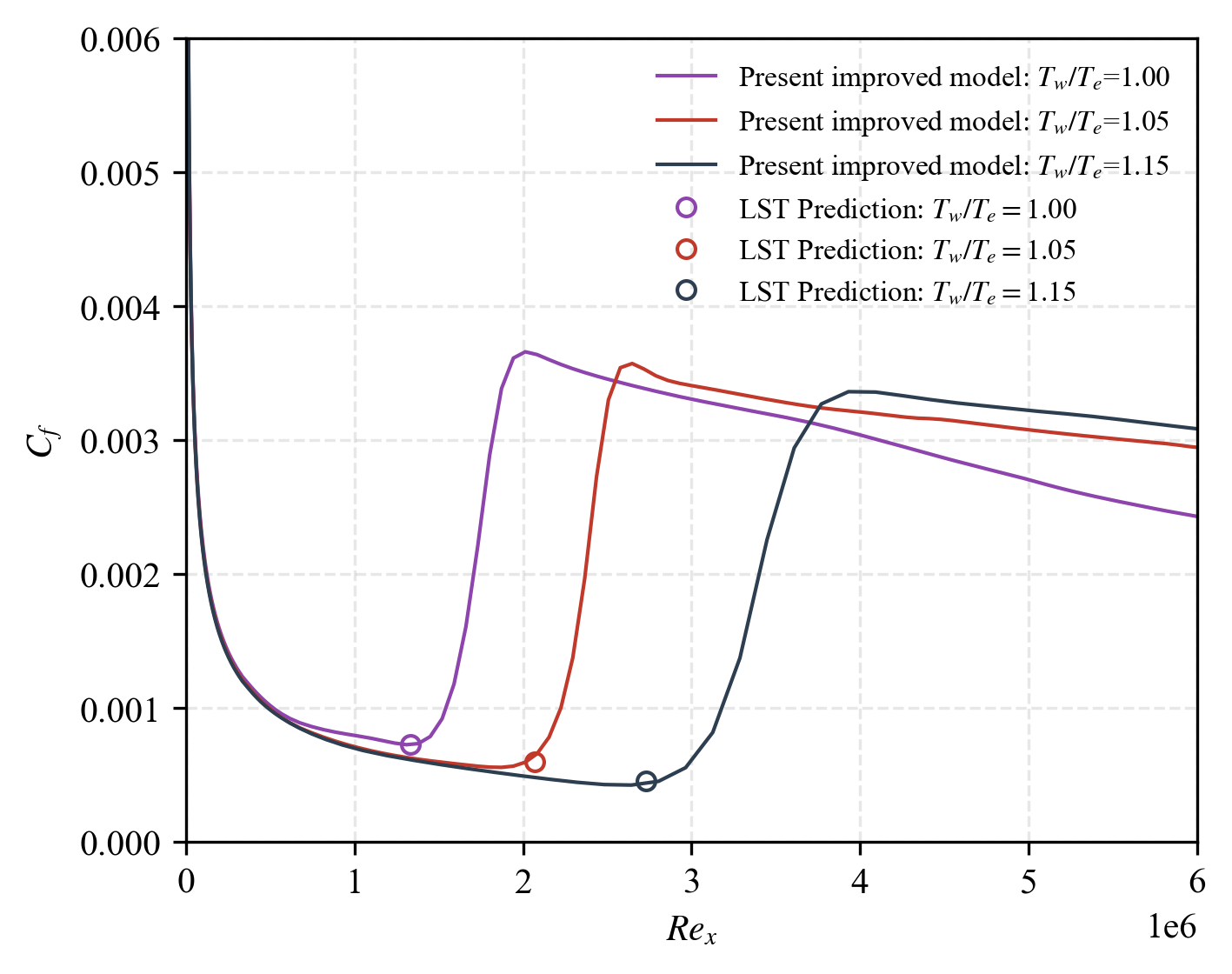}
  \caption{Comparison of skin friction among the improved model and LST results at the cooled wall temperature.}
  \label{f:SKComparehot}
\end{figure}
\begin{figure}[h!]
  \centering
  \includegraphics[width=0.5\textwidth]{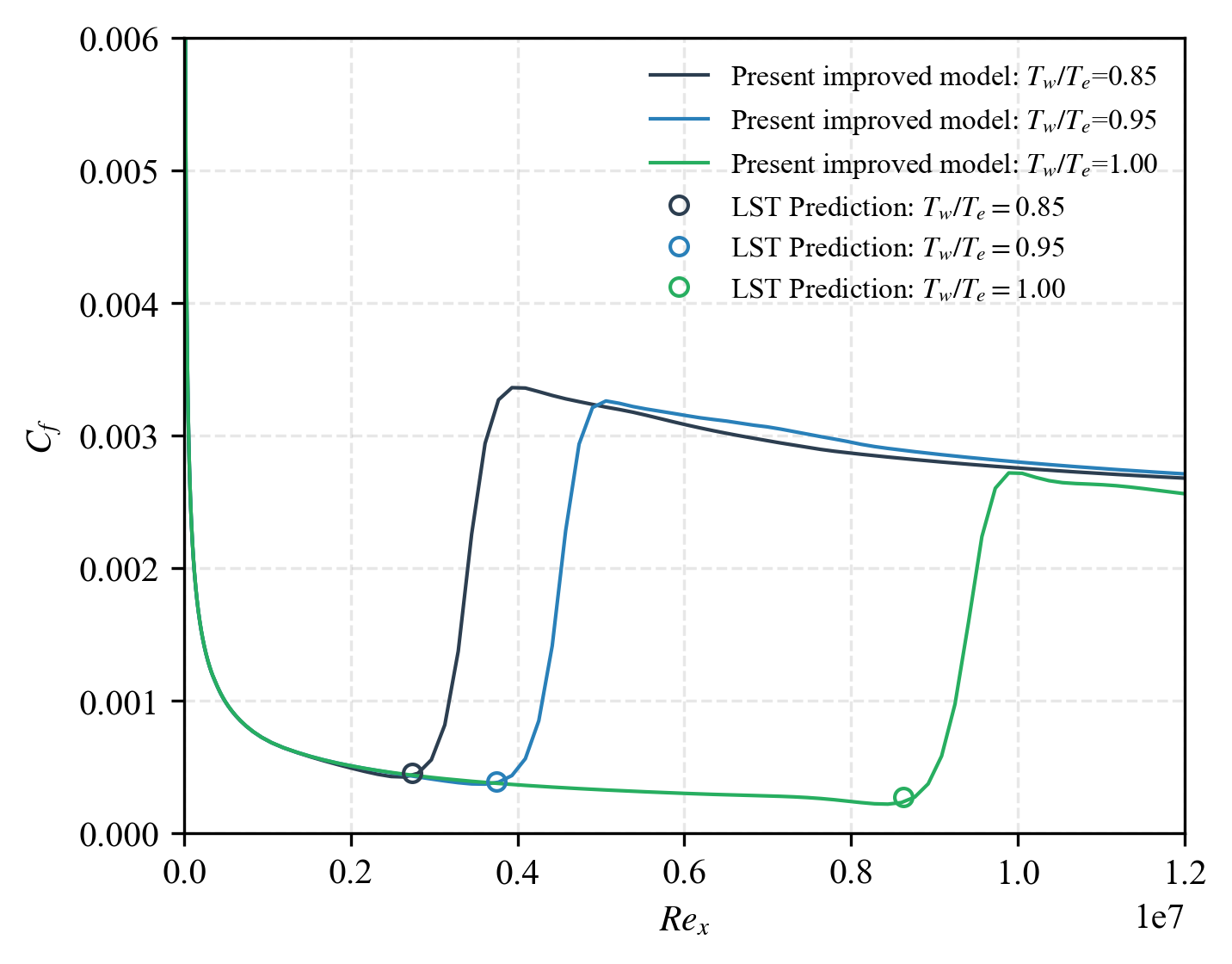}
  \caption{Comparison of skin friction among the improved model and LST results at the heated wall temperature.}
  \label{f:SKCompareCool}
\end{figure}

We present the comparison results between present improved model, LST results (starting points), and experimental result at the adiabatic wall temperature in Fig.~\ref{f:SKCompare}.
The skin friction coefficient ($C_f$) is used to determine the transition location, which is indicated by a sharp rise in $C_f$.
The horizontal axis represents the local Reynolds number.
The solid lines present improved model, the circle indicates the starting transition point of the LST-based results, and the squares indicate the experimental results.
For the LST results, the transition location is determined by the $e^N$ method with $N_\mathrm{crit} = 8.14$.
It can be seen that the improved model and LST result accurately capture the transition location compared to the experimental results, and the prediction error ($\Delta x_\text{tr}/c$) is lower than 5\%.
The results with adiabatic conditions and wall heating ($T_w/T_e = 1.05$ and $T_w/T_e = 1.15$) are shown in Fig.~\ref{f:SKComparehot}.
The improved model matches well LST results at wall heating conditions.
The transition location moves upstream as the wall temperature increases, indicating that wall heating has a significant destabilizing effect on the boundary layer.
Specifically, at $T_w/T_e = 1.15$, the laminar flow region is reduced by approximately half.
The results for the wall-cooling cases ($T_w/T_e = 0.85$ and $T_w/T_e = 0.75$) are presented in Fig.~\ref{f:SKCompareCool}.
The improved model also demonstrates excellent agreement with LST predictions under these cooling conditions.
For $T_w/T_e = 0.85$, the transition location shifts significantly downstream, confirming the strong stabilizing effect of wall cooling on the boundary layer.
In summary, the improved model accurately captures the sensitivity of the transition location to wall temperature variations.
Compared with LST results, the transition location prediction errors ($\Delta x_\text{tr}/c$) of the improved model are within 5\% for all cases.

\begin{figure}[h!]
	\centering
	\subfigure[Velocity ${u/u_e}$\label{f:BLV}]
	{\includegraphics[width=2.8in]{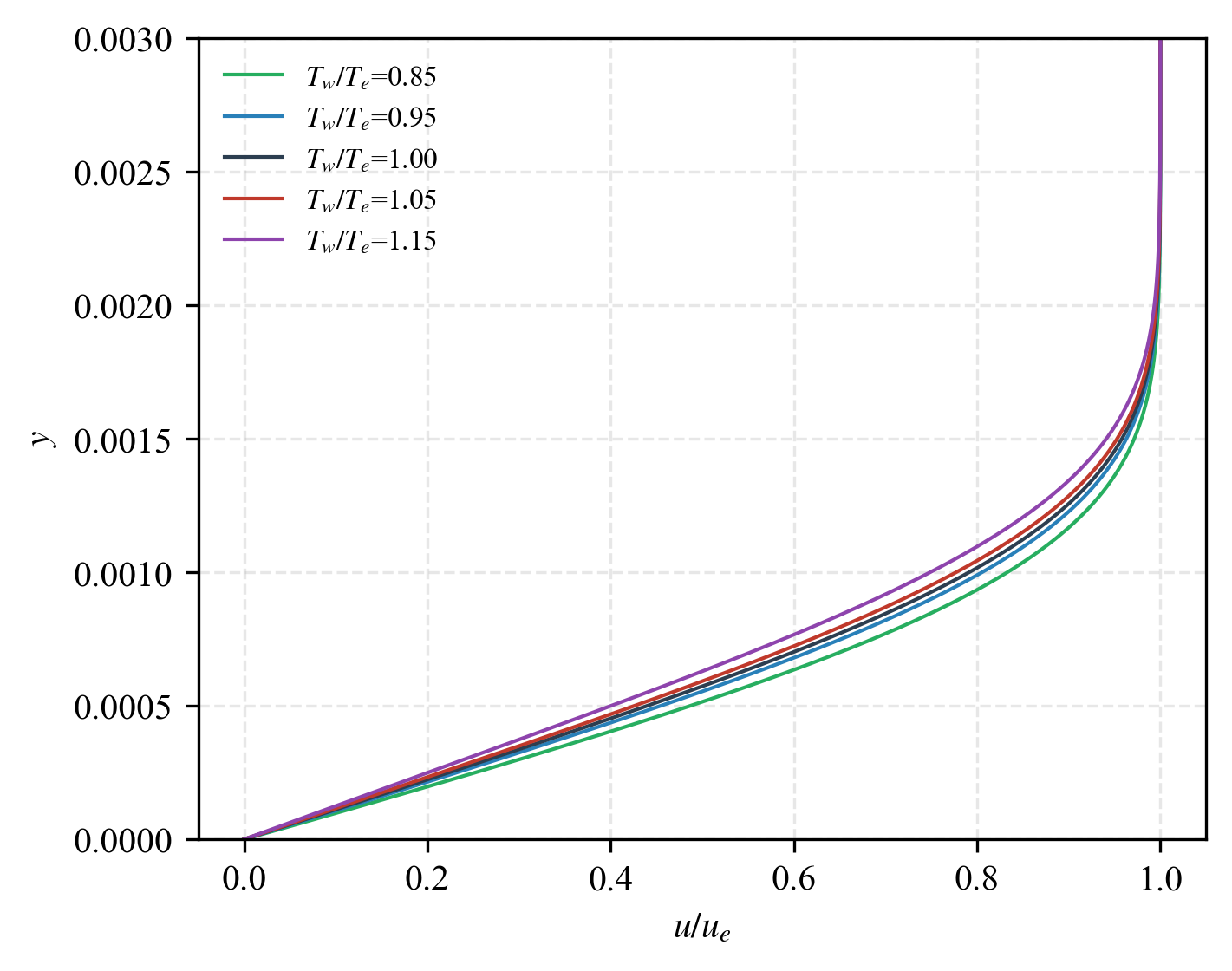}} 
	\subfigure[Temperature ${T/T_e}$\label{f:BLT}]
	{\includegraphics[width=2.8in]{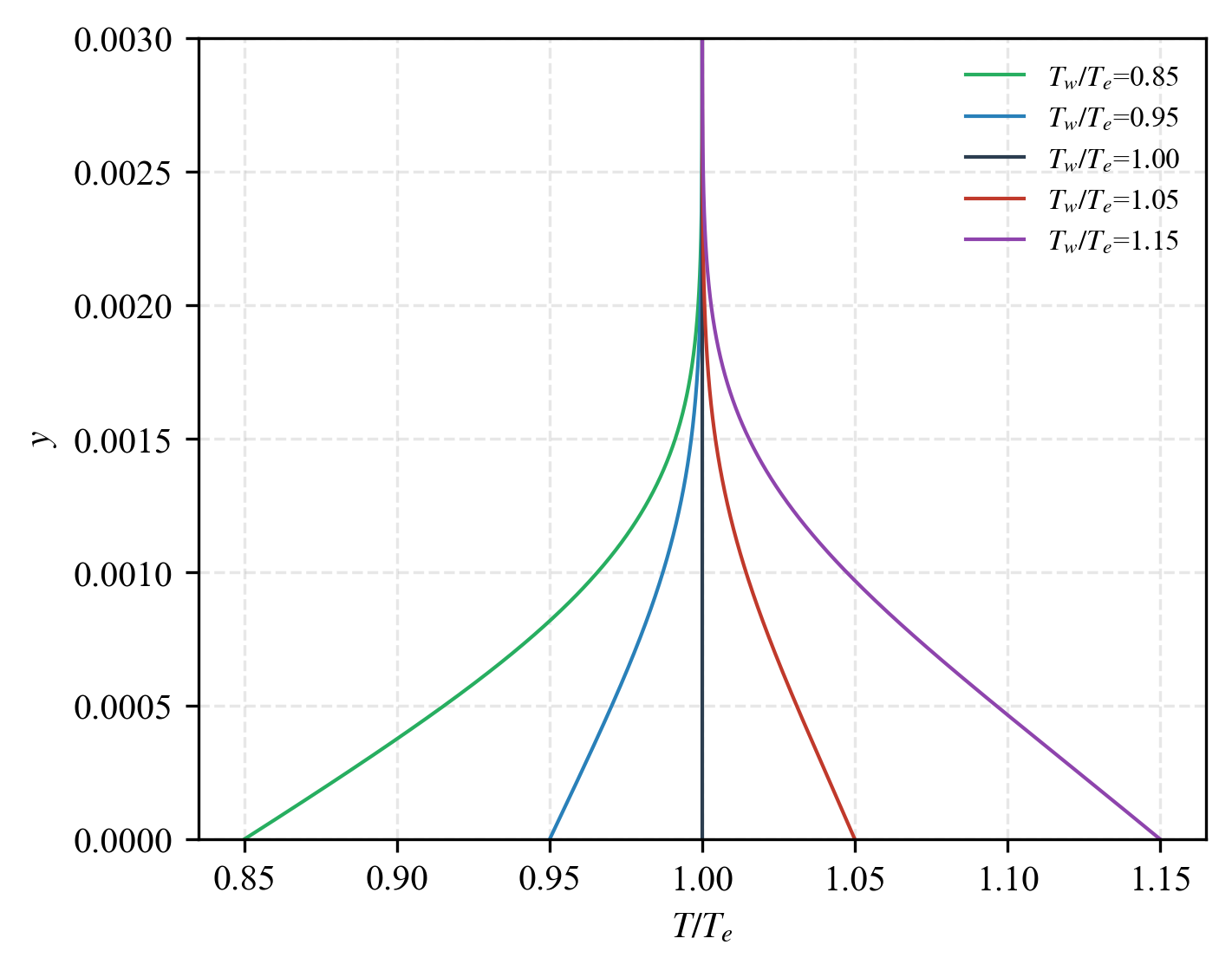}} 
	\caption{The boundary layer velocity and temperature (non-dimensional) comparisons at different temperature difference ratios.}
	\label{f:BLVandT}
\end{figure}

To further illustrate the effects of wall heating and cooling on boundary layer characteristics, we present the boundary layer velocity and temperature profiles at different wall-to-ambient temperature ratios in Fig.~\ref{f:BLVandT}.
The velocity profiles are non-dimensionalized by the edge velocity $u_e$, while the temperature profiles are non-dimensionalized by the edge temperature $T_e$.
From Fig.~\ref{f:BLV}, we observe that wall heating leads to a fuller velocity profile with a higher shape factor, indicating a more unstable boundary layer. Conversely, wall cooling results in a more pronounced velocity gradient near the wall, signifying a more stable boundary layer.
Fig.~\ref{f:BLT} shows that wall heating increases the temperature throughout the boundary layer, while wall cooling decreases it. These changes in temperature profiles further influence the viscosity and density distributions, thereby affecting the stability characteristics of the boundary layer.
Overall, the boundary layer profiles corroborate the observed trends in transition locations, highlighting the destabilizing effect of wall heating and the stabilizing effect of wall cooling.

\section{The Wind Tunnel Test and validation of Improved Model for the Heated Airship}
\label{s:Validation}
We conducted experiments on the designed airship in the FL-52 wind tunnel.
Considering that the real airship's surface primarily experiences heating, the focus of the experiments is to measure the LLT changes after heating.
Furthermore, we validated our improved model to test whether it can capture the impact of heating on transition in a real three-dimensional configuration.

\subsection{The Wind Tunnel Test of the Heated Airship}
\label{s:WindTunnel}
\begin{figure}[h!]
	\centering
	\subfigure[The wind tunnel test setup\label{f:WindTunnel1}]
	{\includegraphics[height=2.0in]{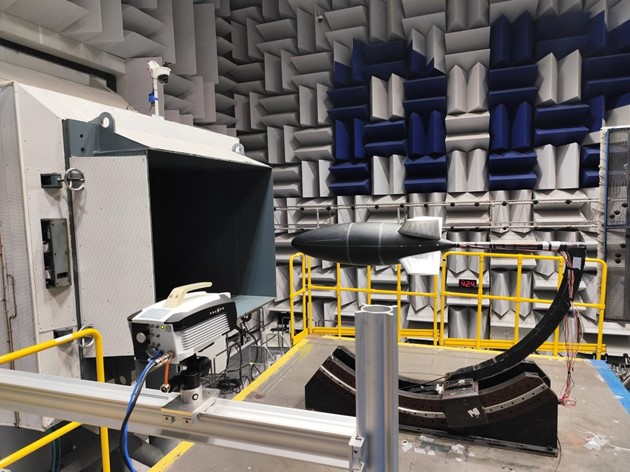}} 
	\subfigure[The airship model\label{f:WindTunnel2}]
	{\includegraphics[height=2.0in]{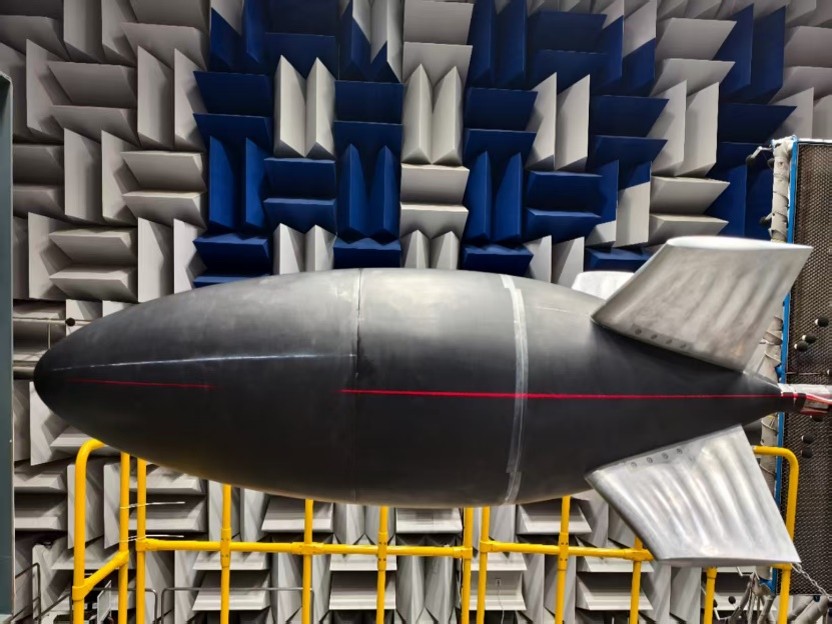}} 
	\caption{The wind tunnel test setup for the airship.}
	\label{f:WindTunnelModel}
\end{figure}

\begin{figure}[h!]
  \centering
  \includegraphics[width=0.7\textwidth]{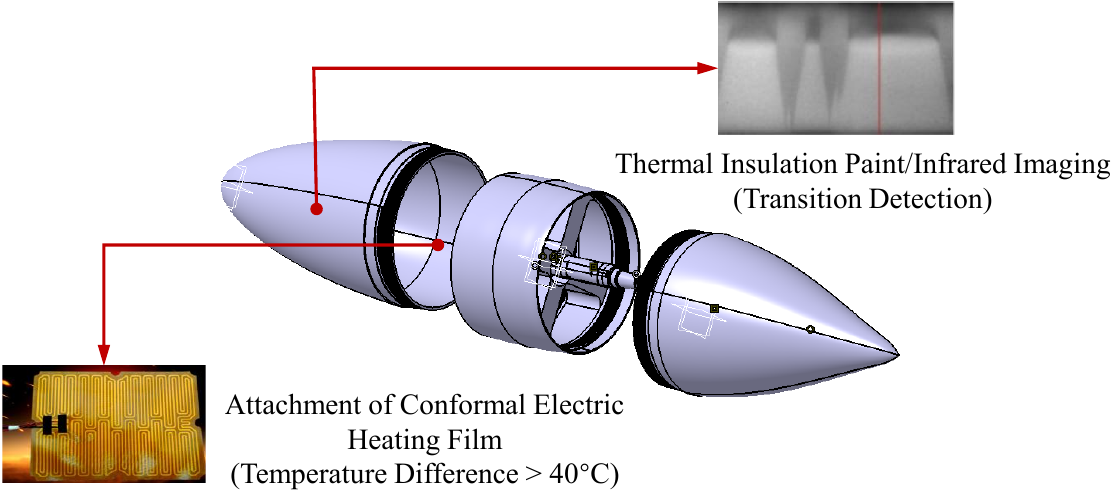}
  \caption{The schematic diagram of heating film and transition detection on the airship model.}
  \label{f:WindTunnelWithHeating}
\end{figure}

The wind tunnel experiments are conducted using the FL-52 wind tunnel open test section platform with a semi-curved blade angle mechanism to support the model, as shown in the Fig.~\ref{f:WindTunnelModel}.
An infrared thermography camera was used to measure the surface temperature, and the transition location was determined based on the temperature gradient.
The tests were conducted at ambient temperature ($T_e$ 288.15K) and heating conditions.
In the heating tests, the model was first heated to 343.34K and then to 371.29K at 50m/s, and to 345.92K and 366.35 at 75m/s using heating films inside the model. The wind tunnel experiments were then conducted, as shown in Fig.~\ref{f:WindTunnelWithHeating}.
The reference length of the model is 1.4$m$, and the reference chord length is 0.2$m^2$. The test angle of attack are 0$^{\circ}$, and the test speeds include 50$m/s$ and 75$m/s$.
The turbulence intensity of the wind tunnel is 0.08\%.

\begin{figure}[h!]
	\centering
	\subfigure[$T_w$ = 288.15 K ($T_w/T_e = 1.0$)\label{f:T20}]
	{\includegraphics[height=1.8in]{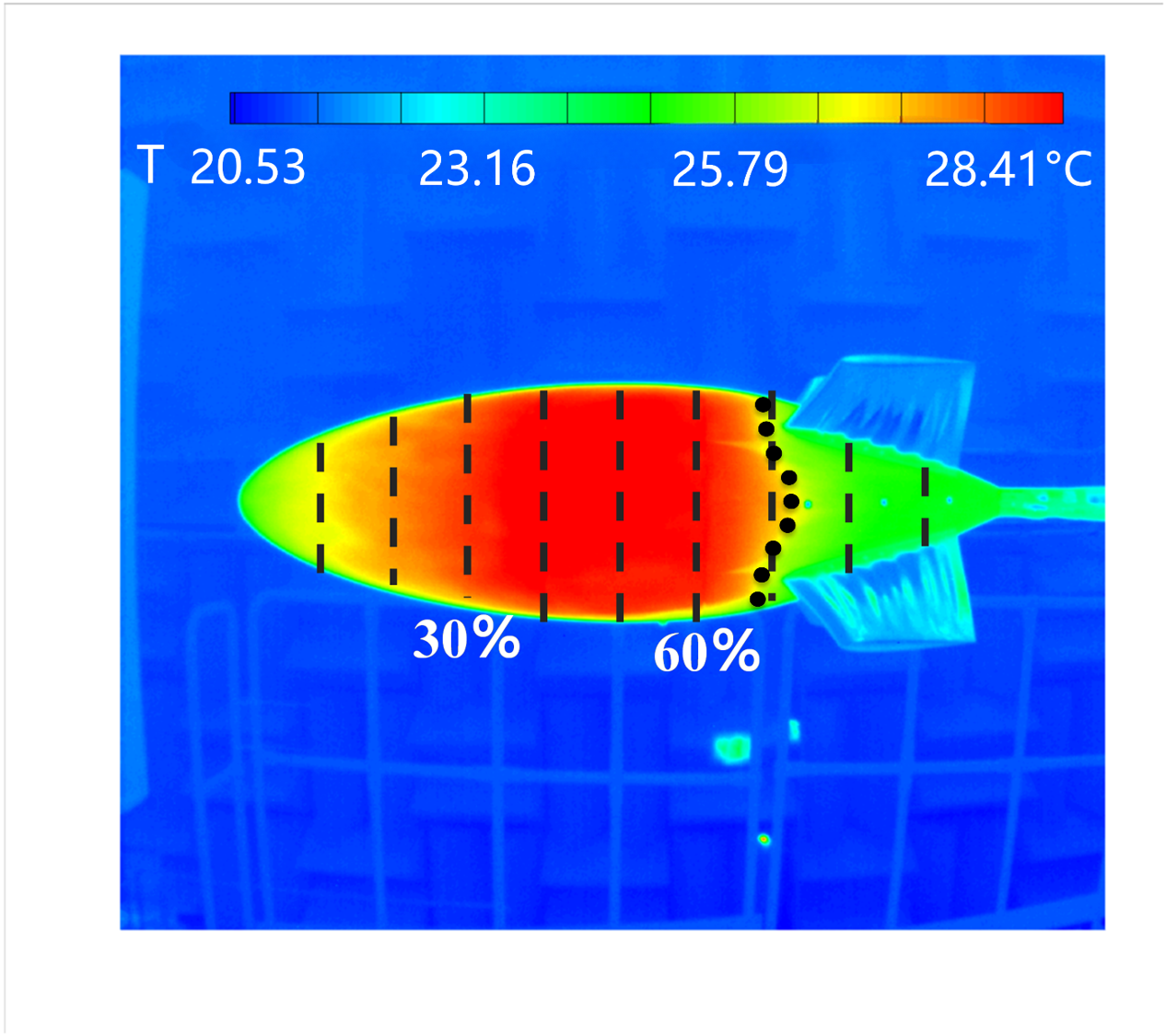}} 
	\subfigure[$T_w$ = 343.34 K ($T_w/T_e = 1.19$)\label{f:T70}]
	{\includegraphics[height=1.8in]{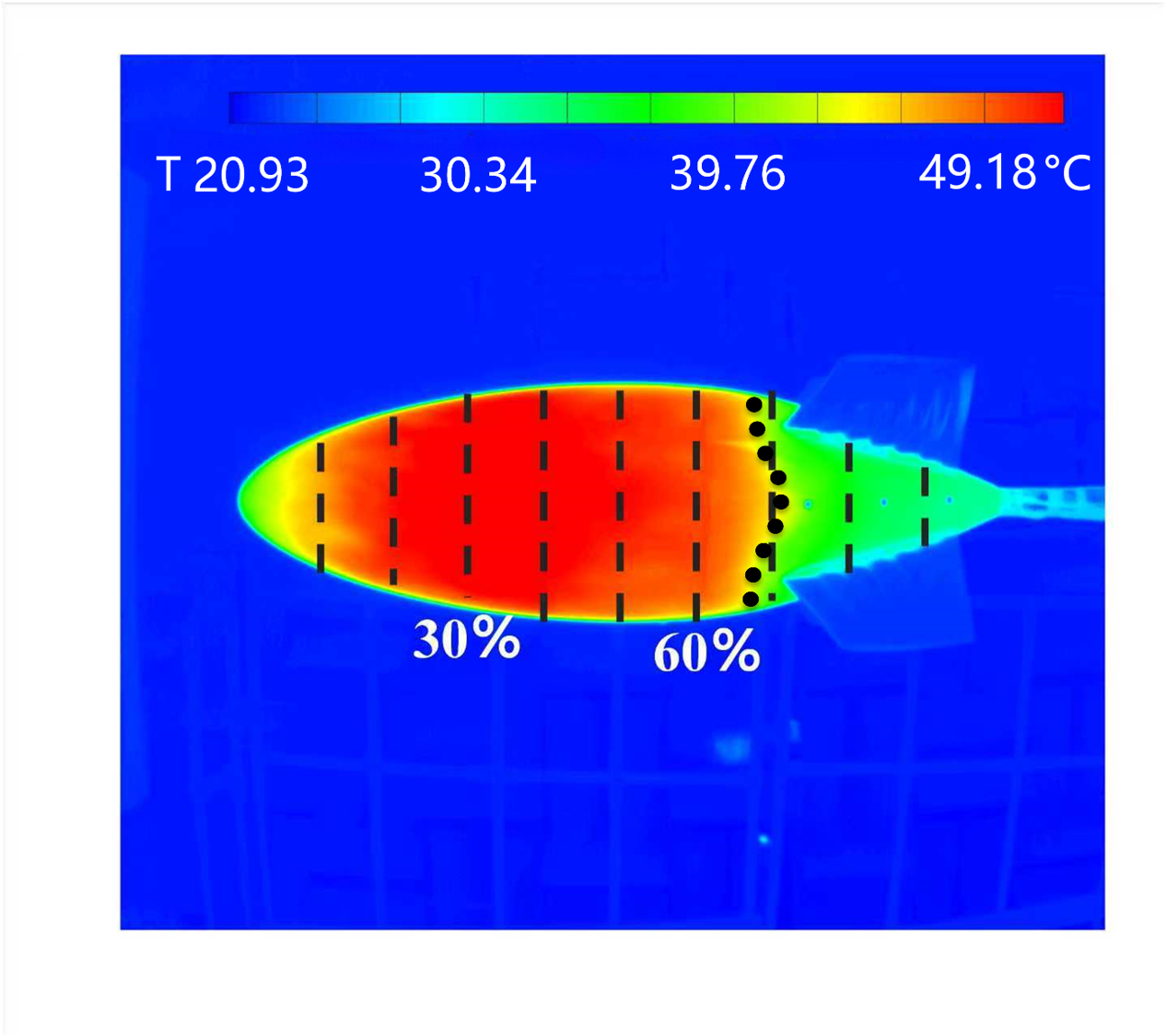}} 
	\subfigure[$T_w$ = 371.29 K ($T_w/T_e = 1.29$)\label{f:T100}]
	{\includegraphics[height=1.8in]{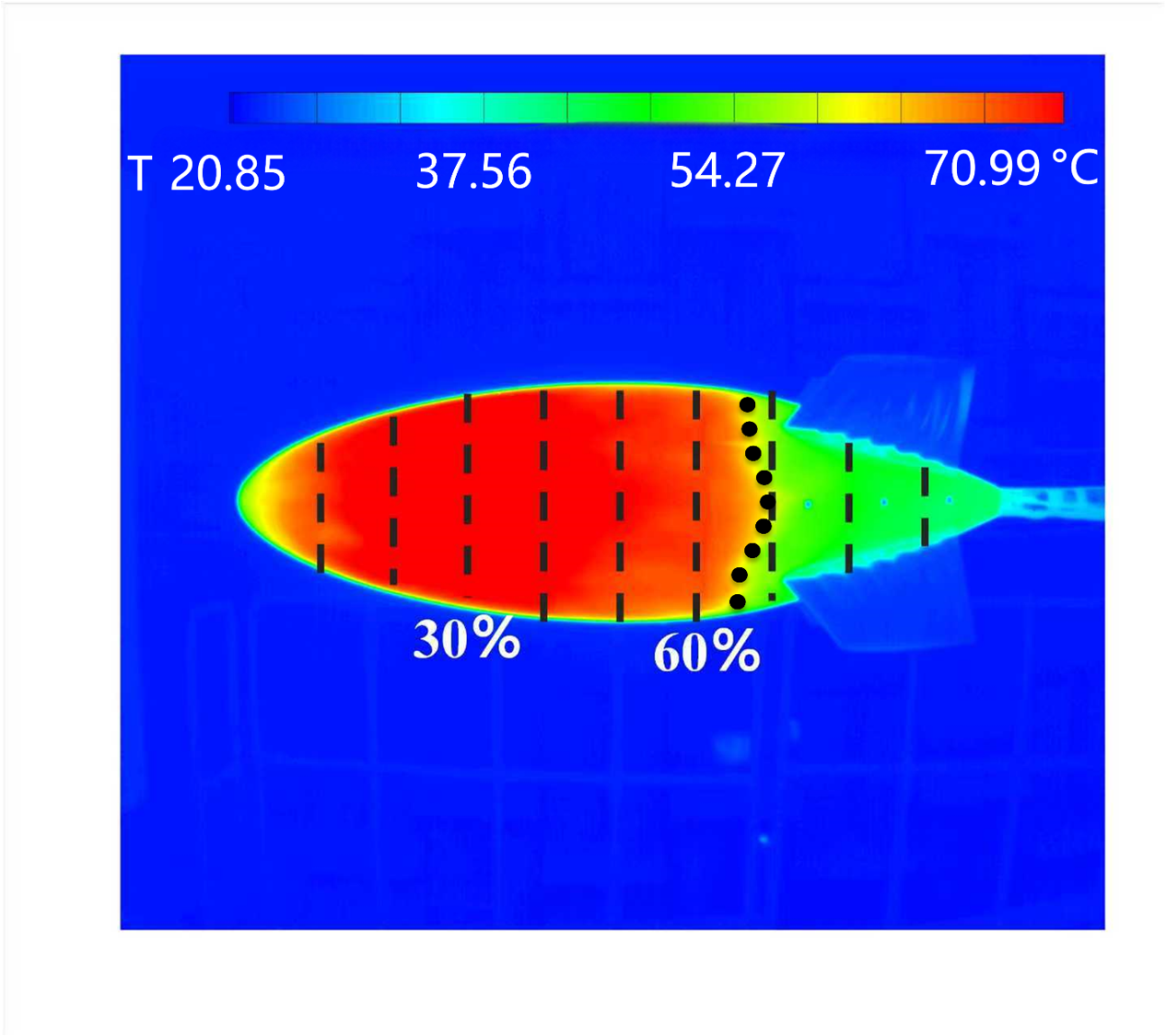}} 
	\caption{The infrared temperature results at $\bm{V = 50m/s}$.}
	\label{f:ExperimentalResultsV50}
\end{figure}

\begin{figure}[h!]
	\centering
	\subfigure[$T_w$ = 288.15 K ($T_w/T_e = 1.0$)\label{f:T20V75}]
	{\includegraphics[height=1.8in]{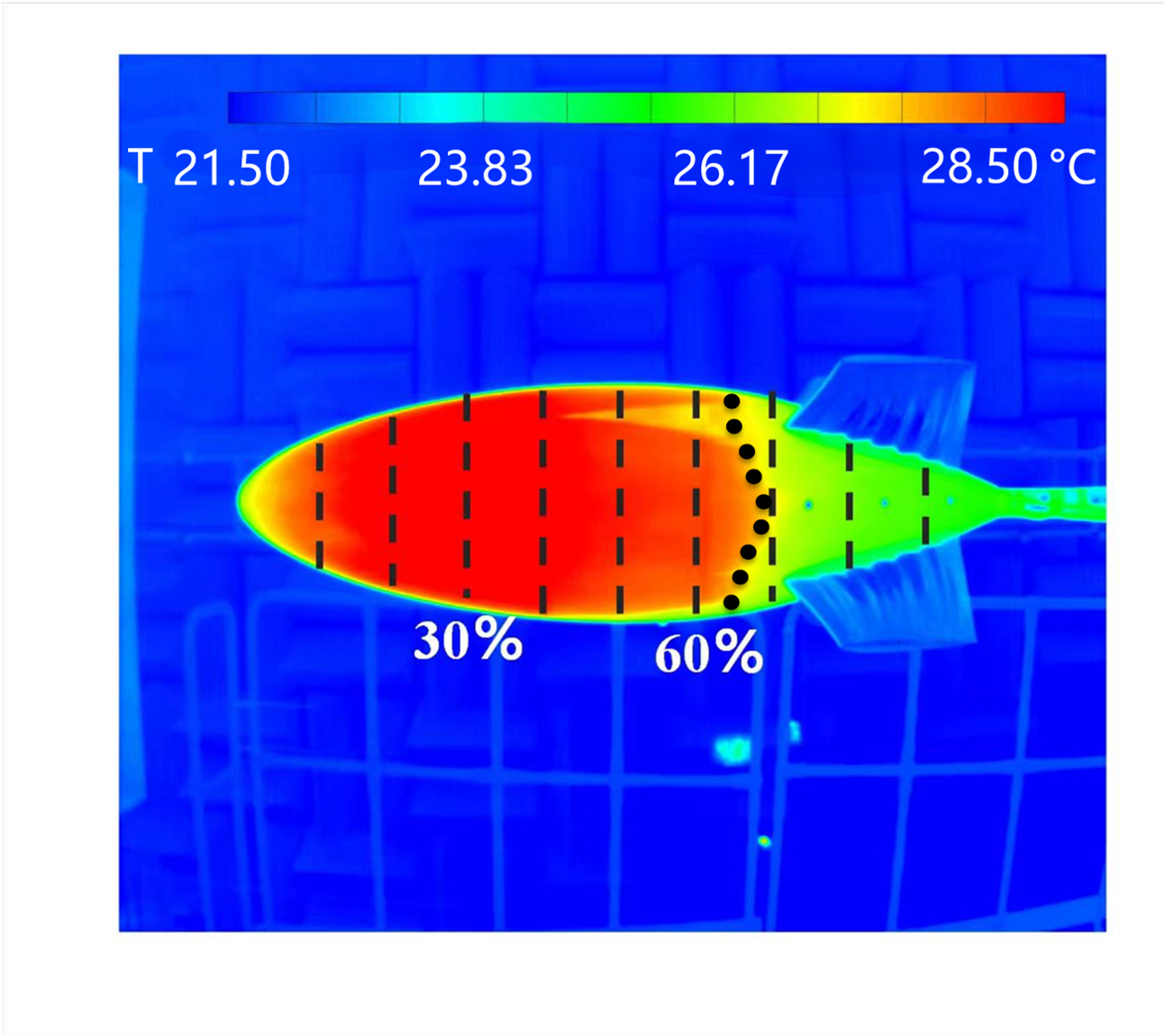}} 
	\subfigure[$T_w$ = 345.92 K ($T_w/T_e = 1.19$)\label{f:T70V75}]
	{\includegraphics[height=1.8in]{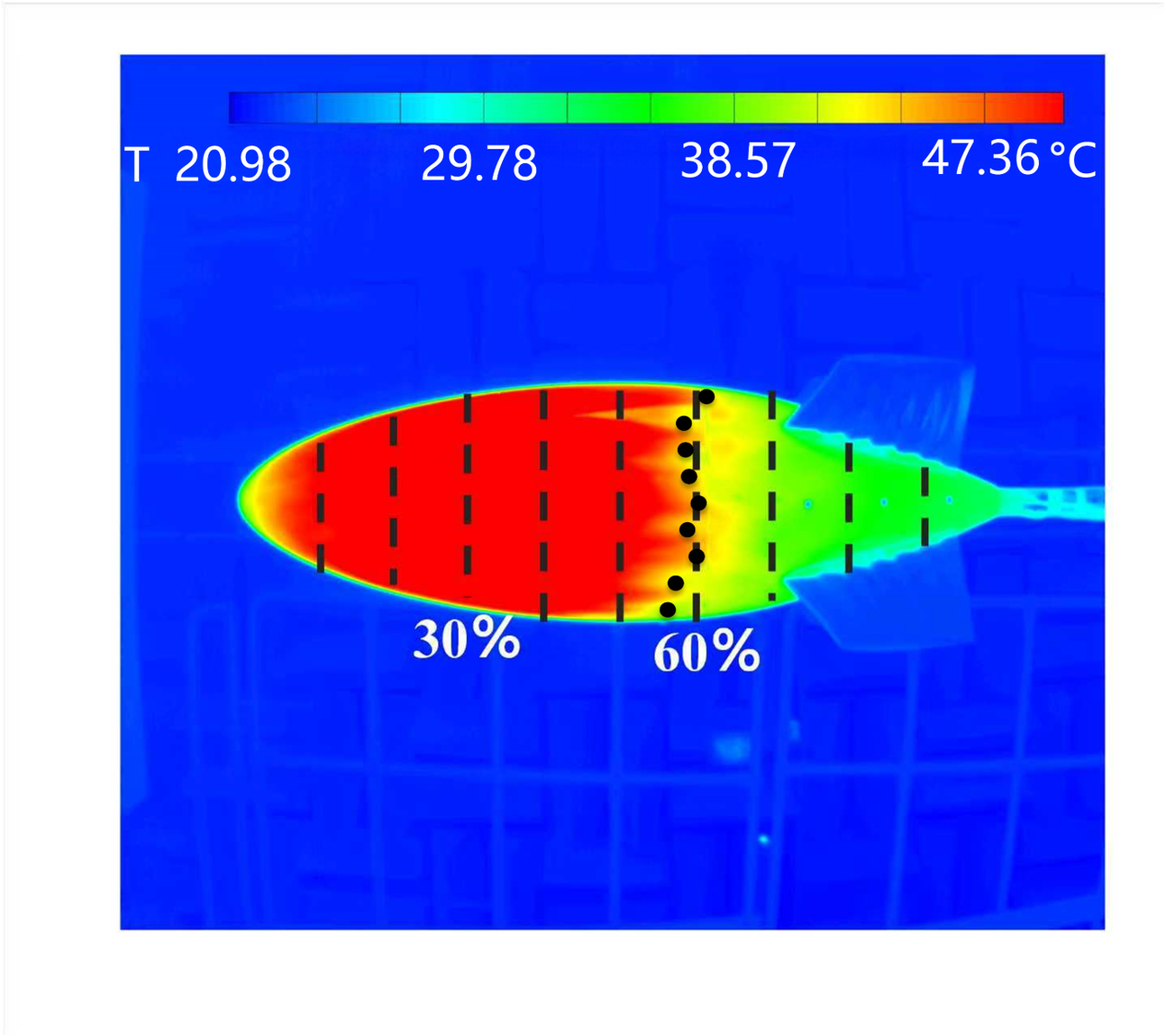}} 
	\subfigure[$T_w$ = 366.35 K ($T_w/T_e = 1.29$)\label{f:T100V75}]
	{\includegraphics[height=1.8in]{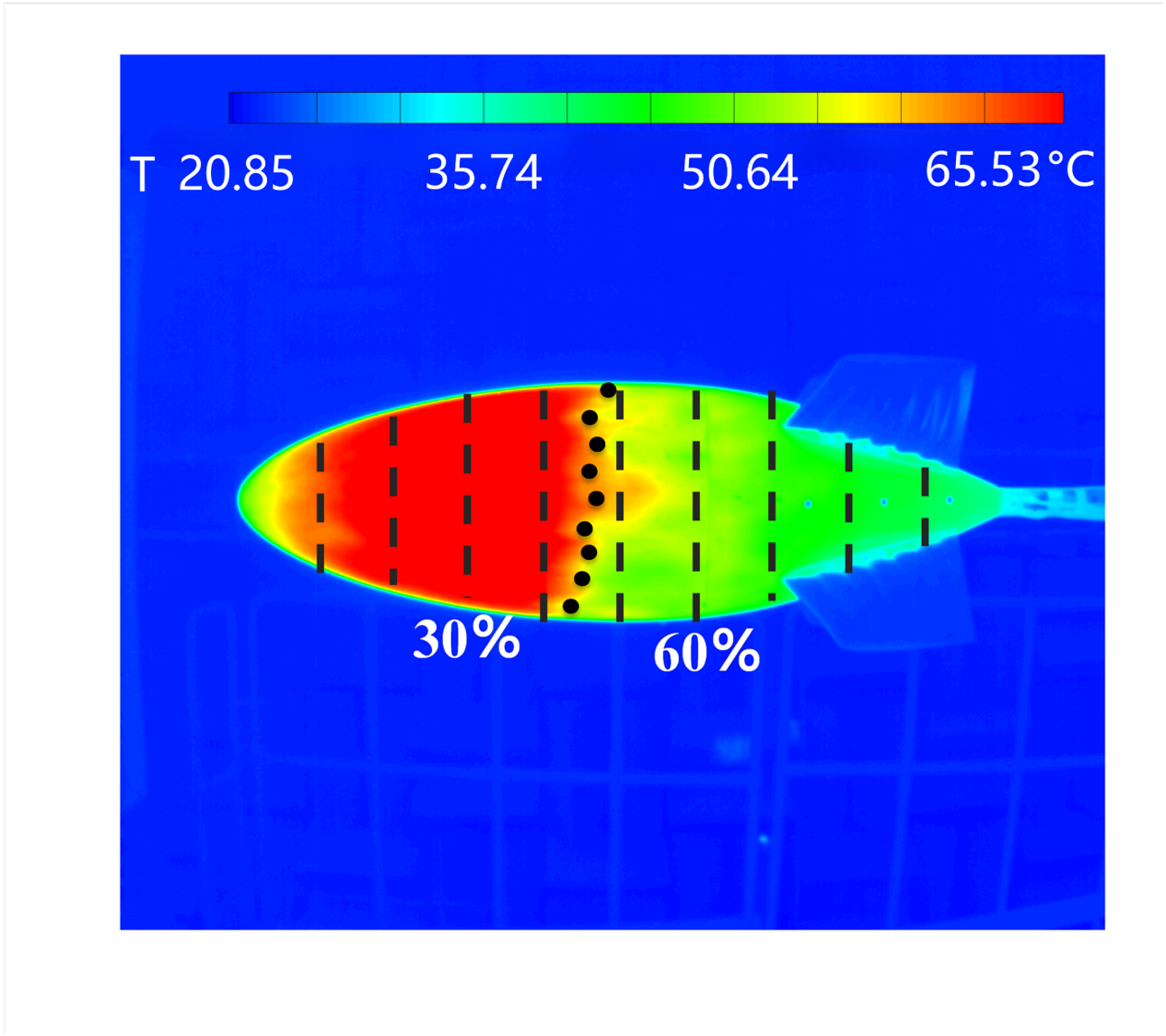}} 
	\caption{The infrared temperature results at $\bm{V = 75m/s}$.}
	\label{f:ExperimentalResultsV75}
\end{figure}

Figures~\ref{f:ExperimentalResultsV50} and \ref{f:ExperimentalResultsV75} present the infrared thermography results across various test conditions.
The experimental methodology relies on the principle that turbulent boundary layers sustain significantly higher convective heat transfer rates than their laminar counterparts.
Following wall heating, the surface is allowed to cool, during which the laminar–turbulent transition front is detected by identifying the abrupt variation in the streamwise surface temperature gradient.
These transition onset points correspond to the distinct steepening of the temperature decay and are explicitly marked with black dots in Figs.~\ref{f:ExperimentalResultsV50} and \ref{f:ExperimentalResultsV75}.
% 这一段不太对

% TODO YS-: 说一下这与前面FSC的结论是一致的
% When the angle of attack is $10^{\circ}$, crossflow is generated. It can be observed that, in the comparison of the images, the onset of transition occurs earlier at the same temperature due to the generation of cross-flow (Fig.~\ref{f:ExperimentalResultsV75A6}).
% Besides, for the crossflow case, heating has also a pronounced effect on the transition location.
% TODO: Add AoA = 6 or AoA = 10, Analyze the influent on TS and CF.

From the infrared temperature results, it can be observed that as the surface temperature of the airship increases, the transition location moves forward.
The results indicate that heating promotes the transition process on the airship surface.
At a velocity of 50 $m/s$ (corresponding to a unit Reynolds number of $3.359\times10^6$), the impact of heating on the transition location is marginal.
However, as the velocity increases to 75 $m/s$ (corresponding to a unit Reynolds number of $5.039\times10^6$), heating exerts a significantly more pronounced effect on the transition location, as shown in Figure~\ref{f:ExperimentalResultsV75}.
This sensitivity is governed by the Reynolds number and the pressure gradient near the transition front. A detailed analysis of the pressure gradient contribution is presented in Section~\ref{s:ValidationImprovedModel}.
At 75 $m/s$, a wall-temperature increase of 57.77 K shifts the transition location upstream by approximately 10\%. When the temperature increase reaches 78.2 K, the transition location advances by about 25\%.
% As analyzed, increasing Reynolds number significantly amplifies the effect of wall heating on the airship, which aligns with the trends observed in Figs.~\ref{f:Retc_vs_lambda_theta} and \ref{f:Retc_surface}. Since the Reynolds numbers encountered in real airship flight are typically high, airships are therefore prone to surface-heating effects during daytime operation, which can degrade the drag-reduction performance. Consequently, the wall-heating effect must be considered.

\subsection{Validation of the Improved Stability-based Model}
\label{s:ValidationImprovedModel}
\begin{figure}[h!]
	\centering
	\subfigure[The model and mesh topology\label{f:Topo}]
	{\includegraphics[height=2.0in]{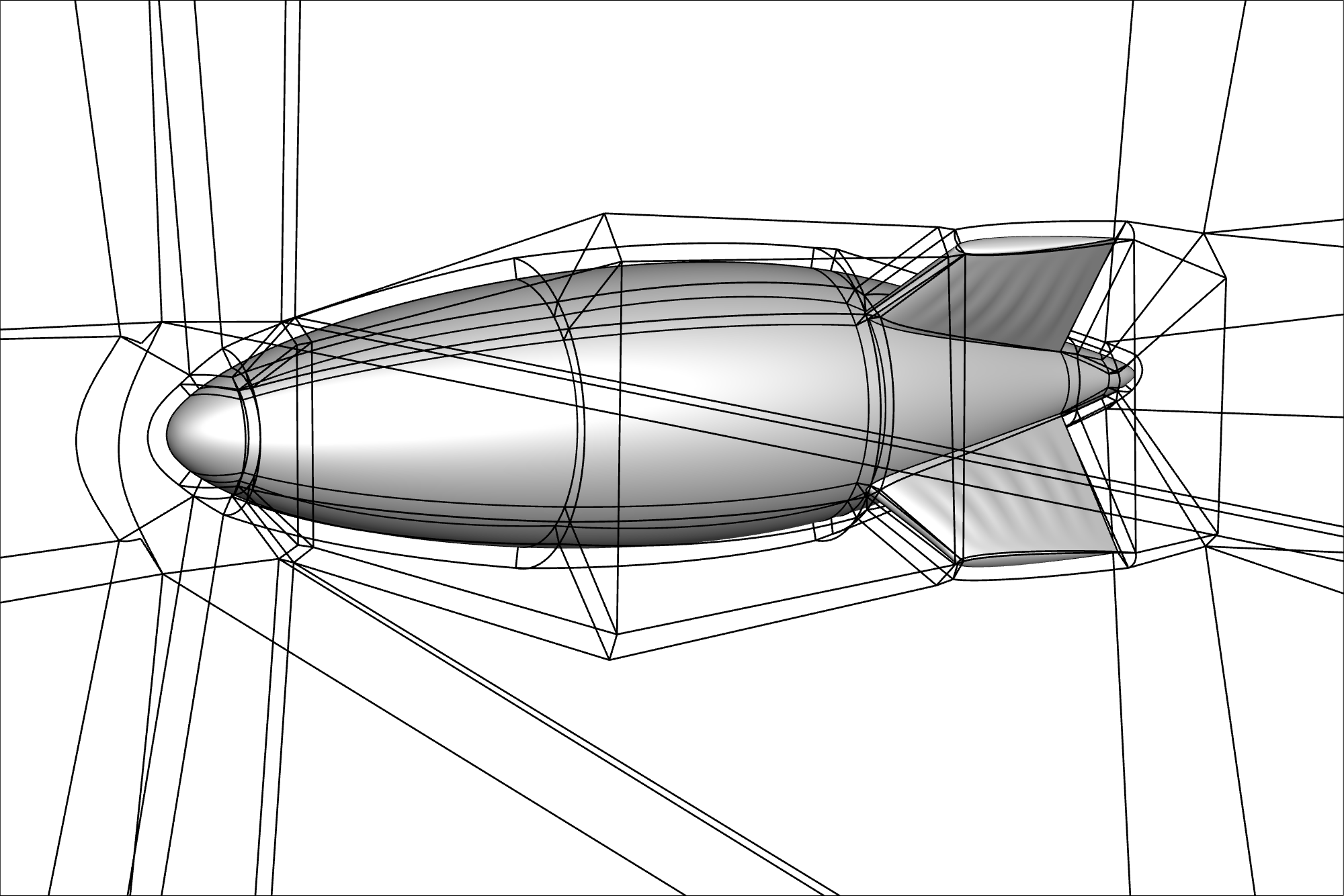}} 
	\subfigure[The surface mesh\label{f:Mesh}]
	{\includegraphics[height=2.0in]{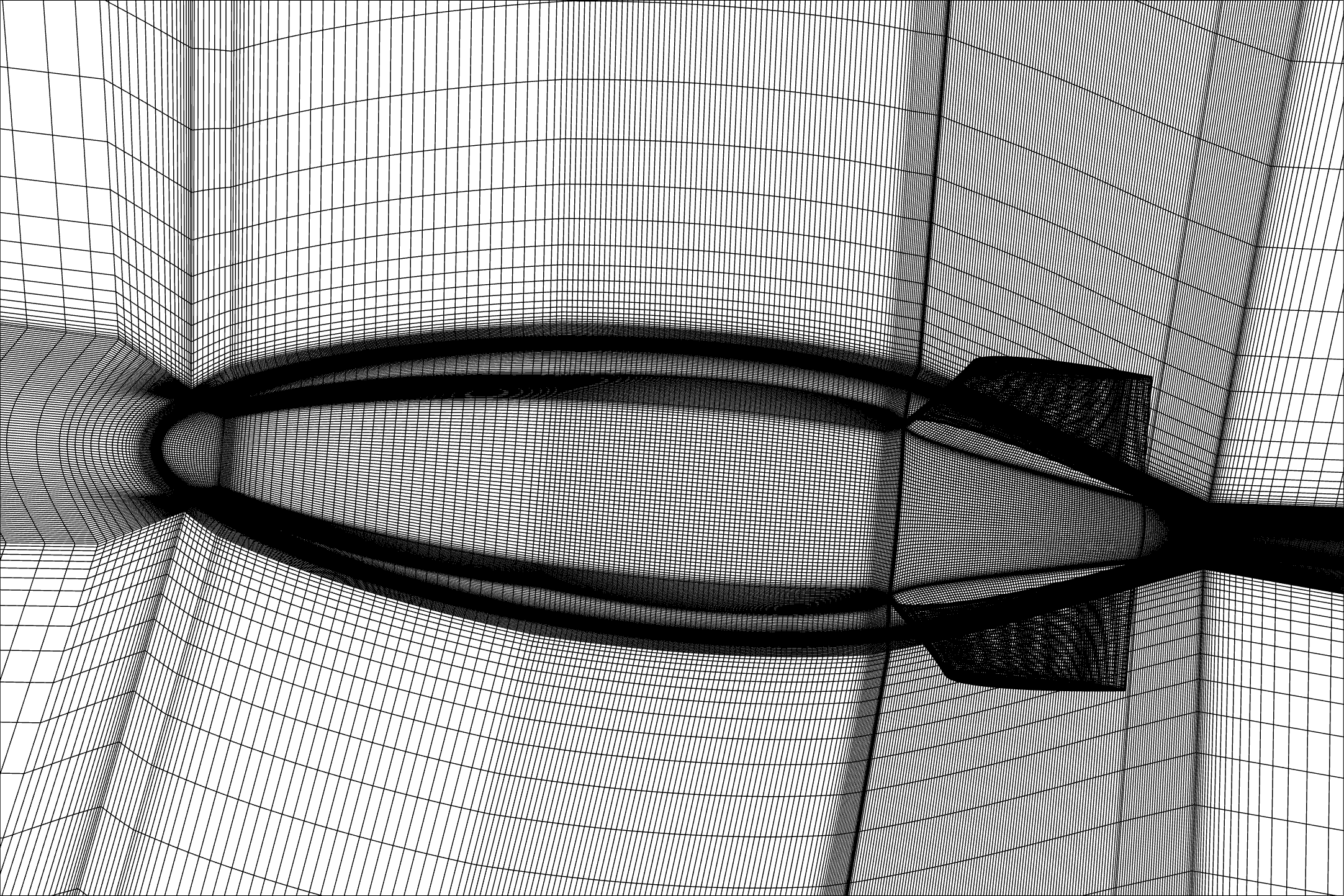}} 
	\caption{The computation mesh for the airship.}
	\label{f:AirShipMesh}
\end{figure}

\begin{figure}[htbp!]
	\centering
	\subfigure[$T_w$ = 288.15 K ($T_w/T_e = 1.0$)\label{f:V50T20Correct}]
	{\includegraphics[height=1.6in]{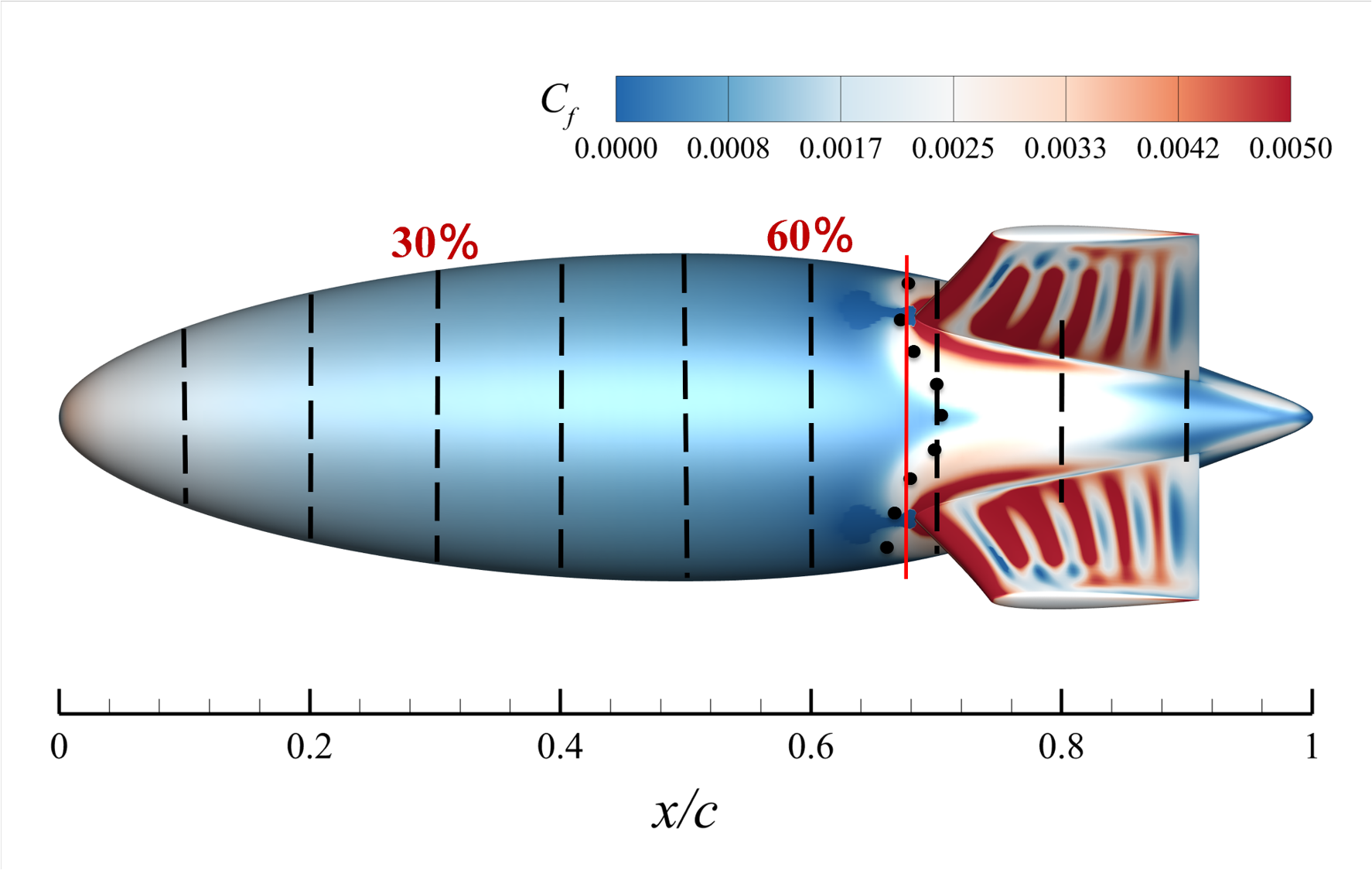}} 
	\subfigure[$T_w$ = 343.34 K ($T_w/T_e = 1.19$)\label{f:V50Correct}]
	{\includegraphics[height=1.6in]{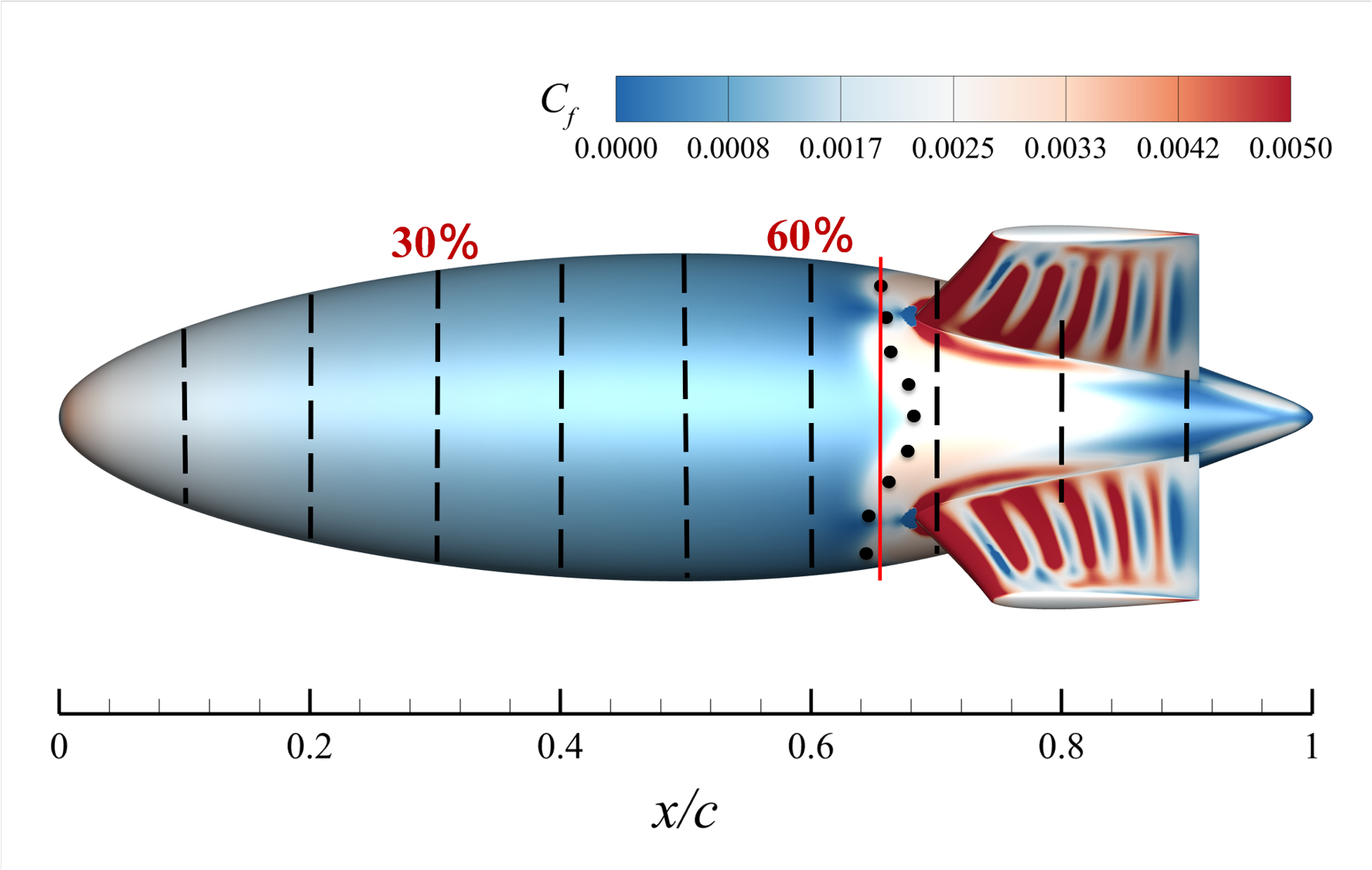}} 
	\subfigure[$T_w$ = 371.29 K ($T_w/T_e = 1.29$)\label{f:V50T100Correct}]
	{\includegraphics[height=1.6in]{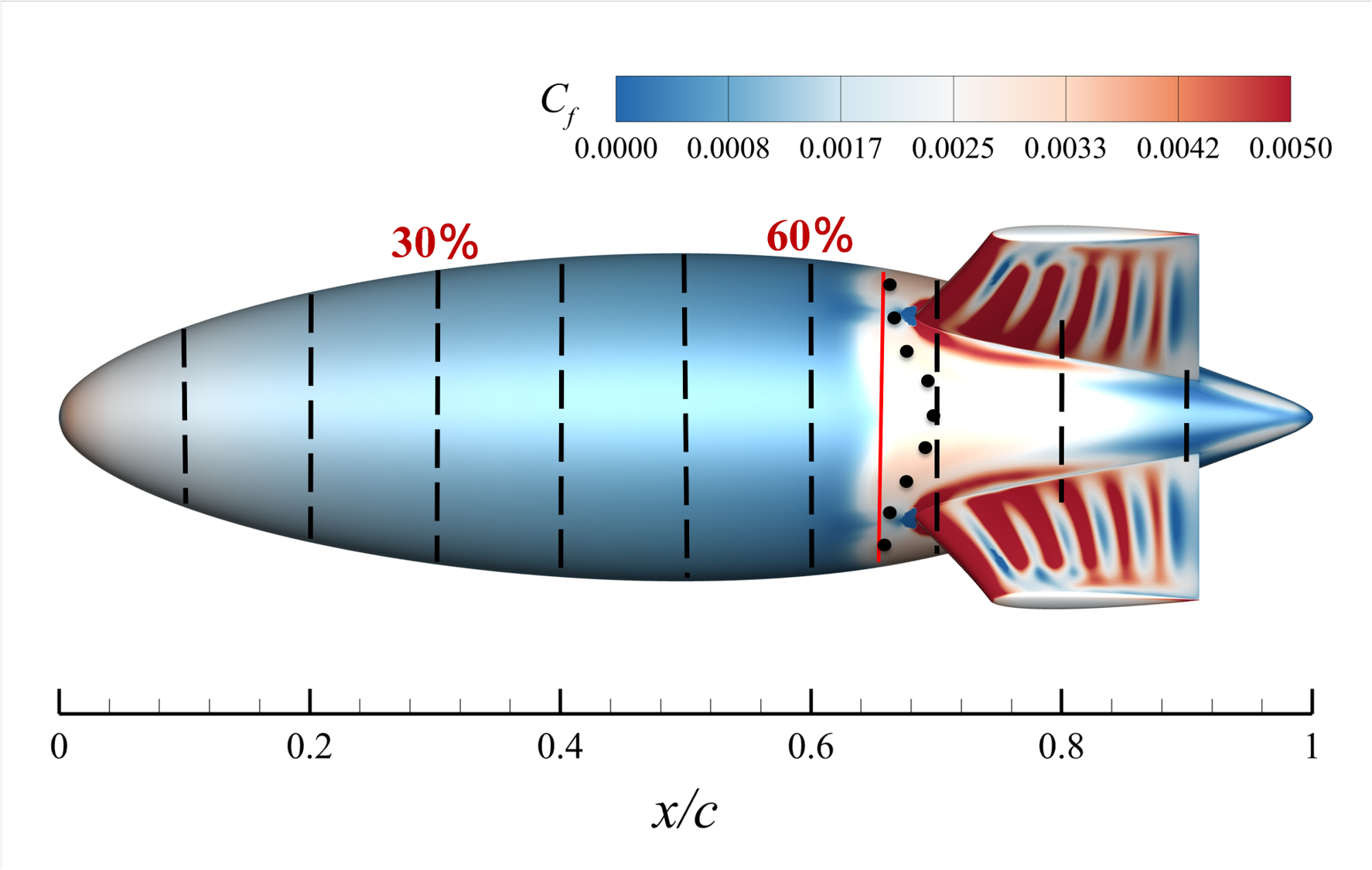}} 
	\caption{The transition locations comparison between present improved model and experimental results at $\bm{V = 50m/s}$.}
	\label{f:AirshipCorrection50}
\end{figure}

\begin{figure}[htbp!]
	\centering
	\subfigure[$T_w$ = 288.15 K ($T_w/T_e = 1.0$)\label{f:V75T20NoCorrect}]
	{\includegraphics[height=1.6in]{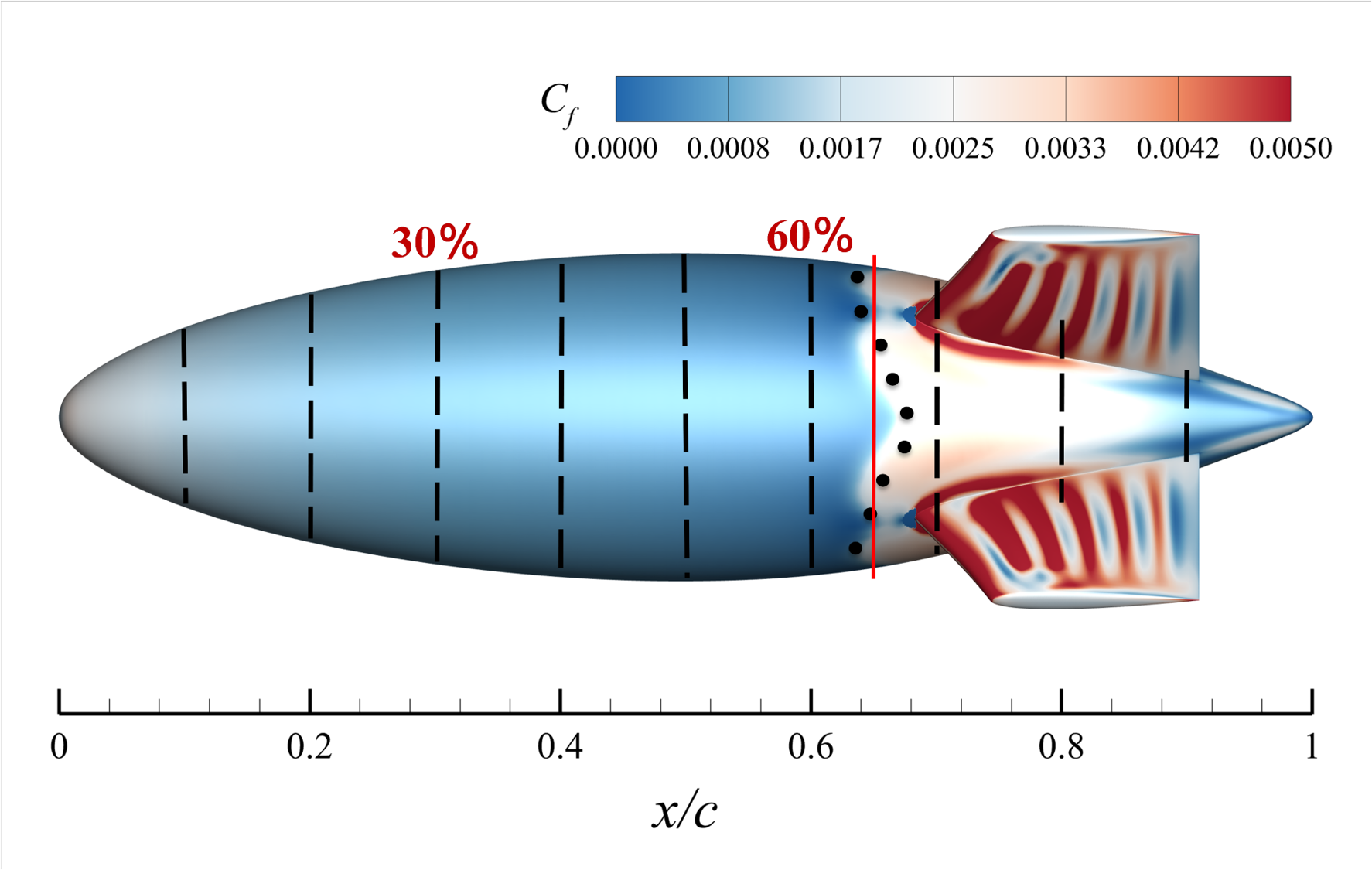}} 
	\subfigure[$T_w$ = 343.34 K ($T_w/T_e = 1.19$)\label{f:V75T70NoCorrect}]
	{\includegraphics[height=1.6in]{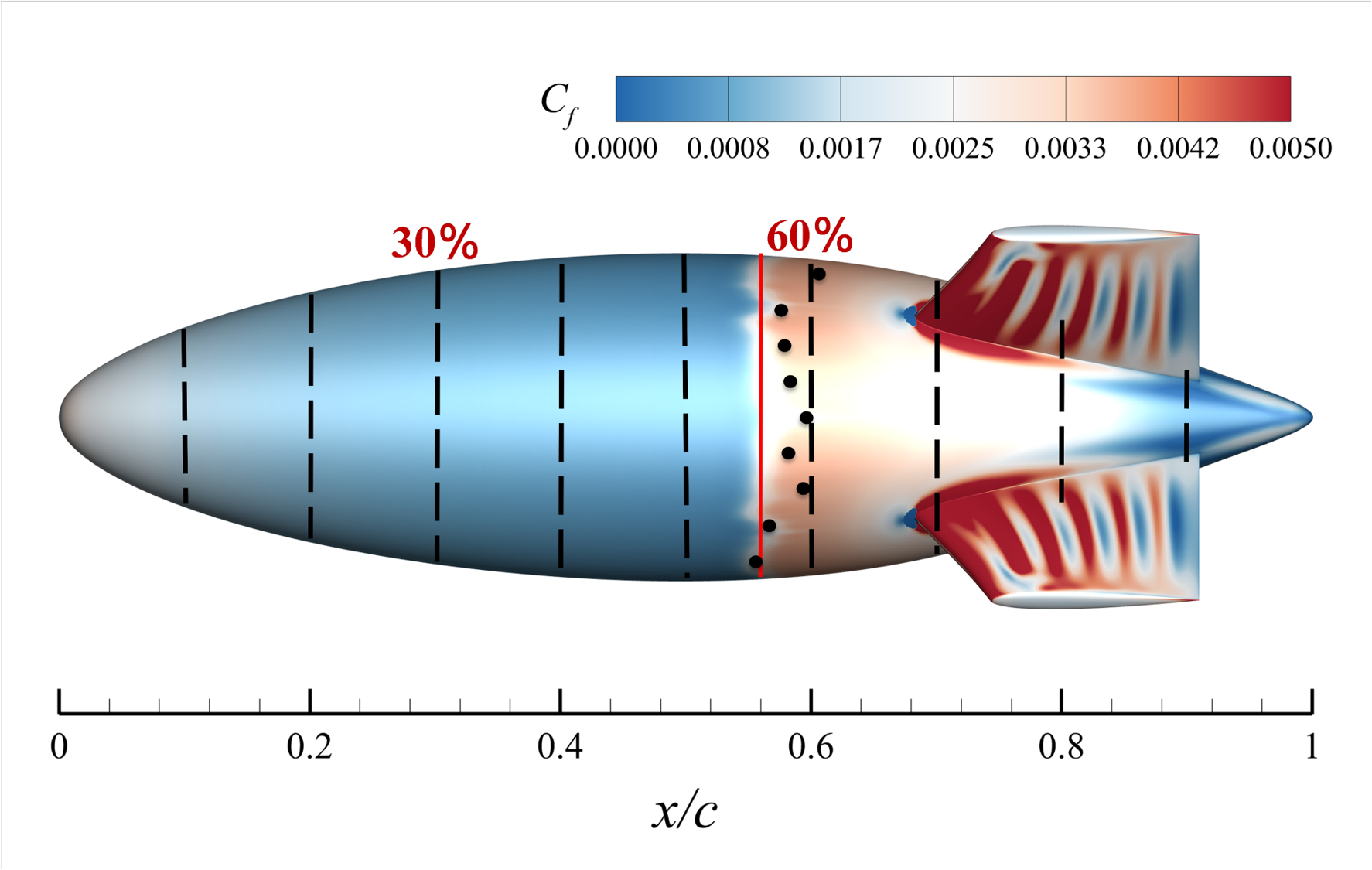}} 
	\subfigure[$T_w$ = 371.29 K ($T_w/T_e = 1.29$)\label{f:V75T100NoCorrect}]
	{\includegraphics[height=1.6in]{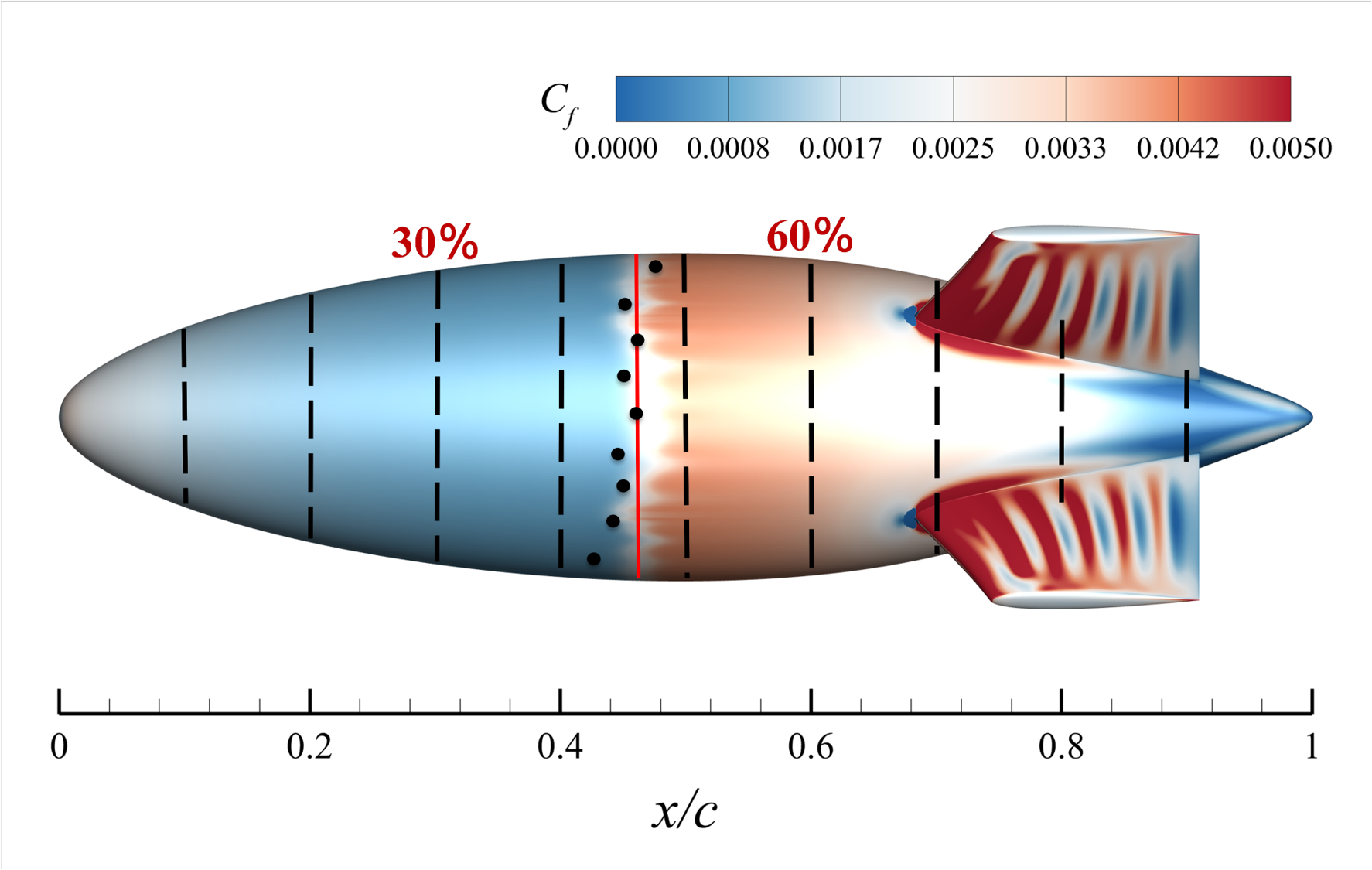}} 
	\caption{The transition locations comparison between present improved model and experimental results at $\bm{V = 75m/s}$.}
	\label{f:AirshipCorrection75}
\end{figure}

\begin{figure}[!t]
	\centering
	\subfigure[$T_w$ = 288.15 K ($T_w/T_e = 1.0$)\label{f:T20NoCorrect}]
	{\includegraphics[height=1.6in]{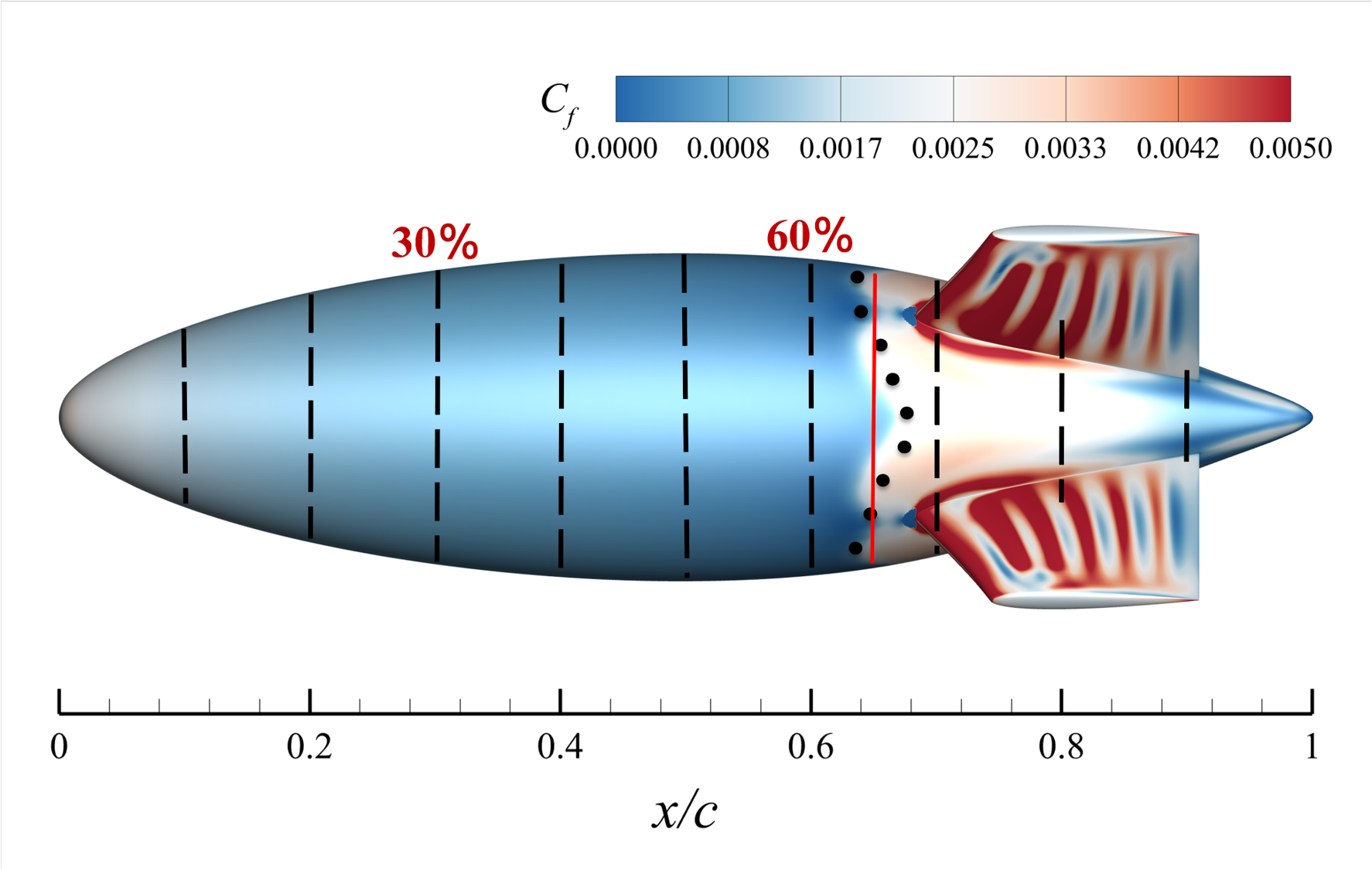}} 
	\subfigure[$T_w$ = 343.34 K ($T_w/T_e = 1.19$)\label{f:T70NoCorrect}]
	{\includegraphics[height=1.6in]{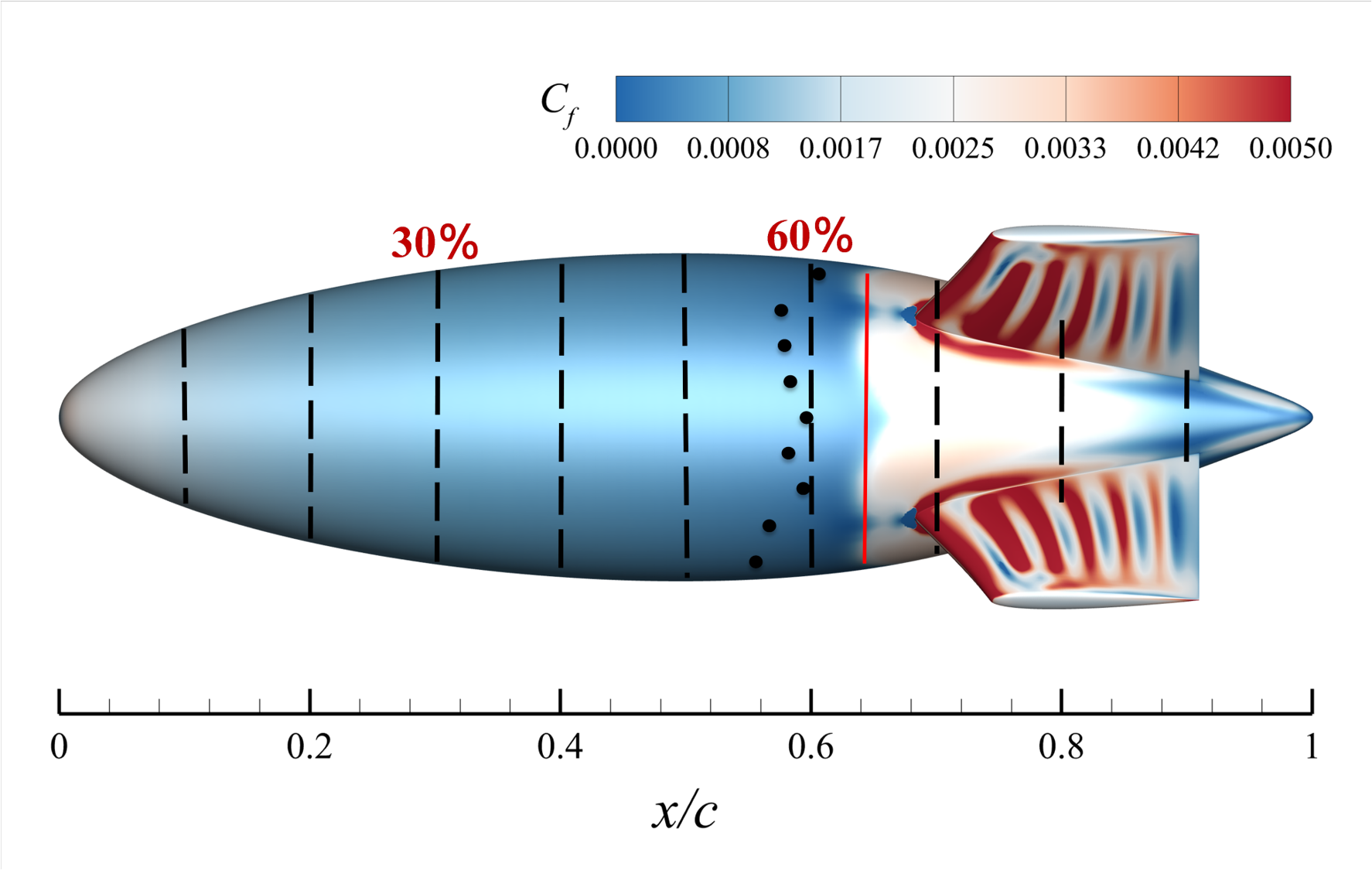}} 
	\subfigure[$T_w$ = 371.29 K ($T_w/T_e = 1.29$)\label{f:T100NoCorrect}]
	{\includegraphics[height=1.6in]{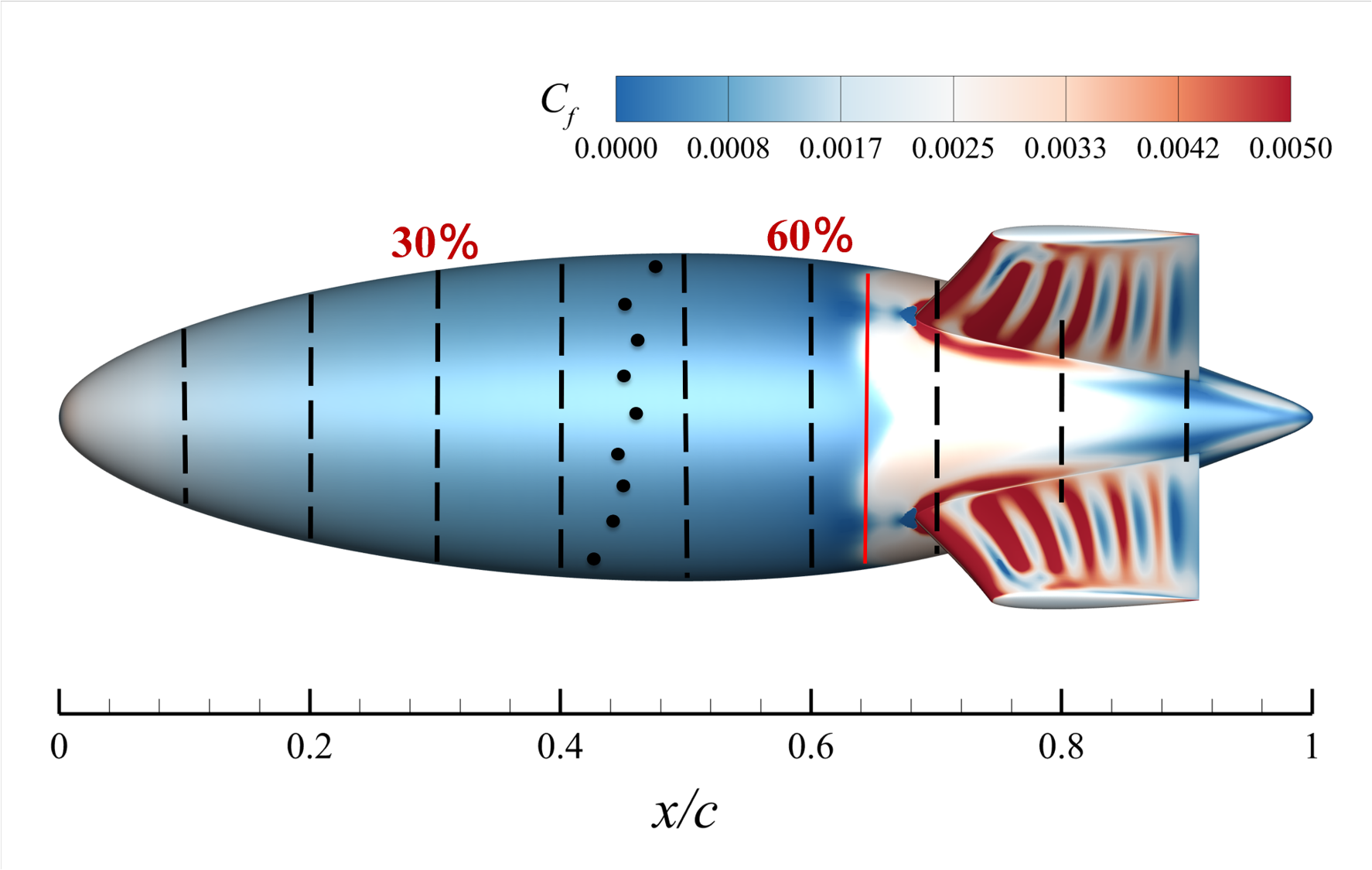}} 
	\caption{The transition locations comparison by using the one-equation model proposed by~\citet{franccois2023simplified} and experimental results at $\bm{V = 75m/s}$.}
	\label{f:AirshipNoCorrection75}
\end{figure}

Based on the experimental results, we validate our improved model. The model and computational mesh are shown in the Fig.~\ref{f:AirShipMesh}. The grid size (half model) is 9,619,200 cells. The first layer in the wall-normal direction has $y^+$=0.8, with a growth rate of 1.15.
The simulation results using the present improved model are shown in Fig.~\ref{f:AirshipCorrection50} and \ref{f:AirshipCorrection75}.
In Fig.~\ref{f:AirshipCorrection50} and \ref{f:AirshipCorrection75}, the solid lines represent the simulation average transition locations at the streamwise direction, while the black solid circles indicate the experimental results.
It can be observed that under different velocities and heating conditions, the computational results agree well with the experimental data. Compared with the experiments, the relative error is between 2\% and 6\%.
We also simulate the intermittency model proposed by~\citet{franccois2023simplified}, the simulation results and experimental comparisons are shown in Fig. \ref{f:AirshipNoCorrection75}.
The transition locations at different wall temperatures are almost identical, which contradicts the experimental results.
This demonstrates the necessity of accounting for the effect of temperature differences, thereby capturing its impact on the transition location and laminar region.

\begin{figure}[htbp!]
	\centering
	\subfigure[50$m/s$\label{f:Cp50}]
	{\includegraphics[height=2.2in]{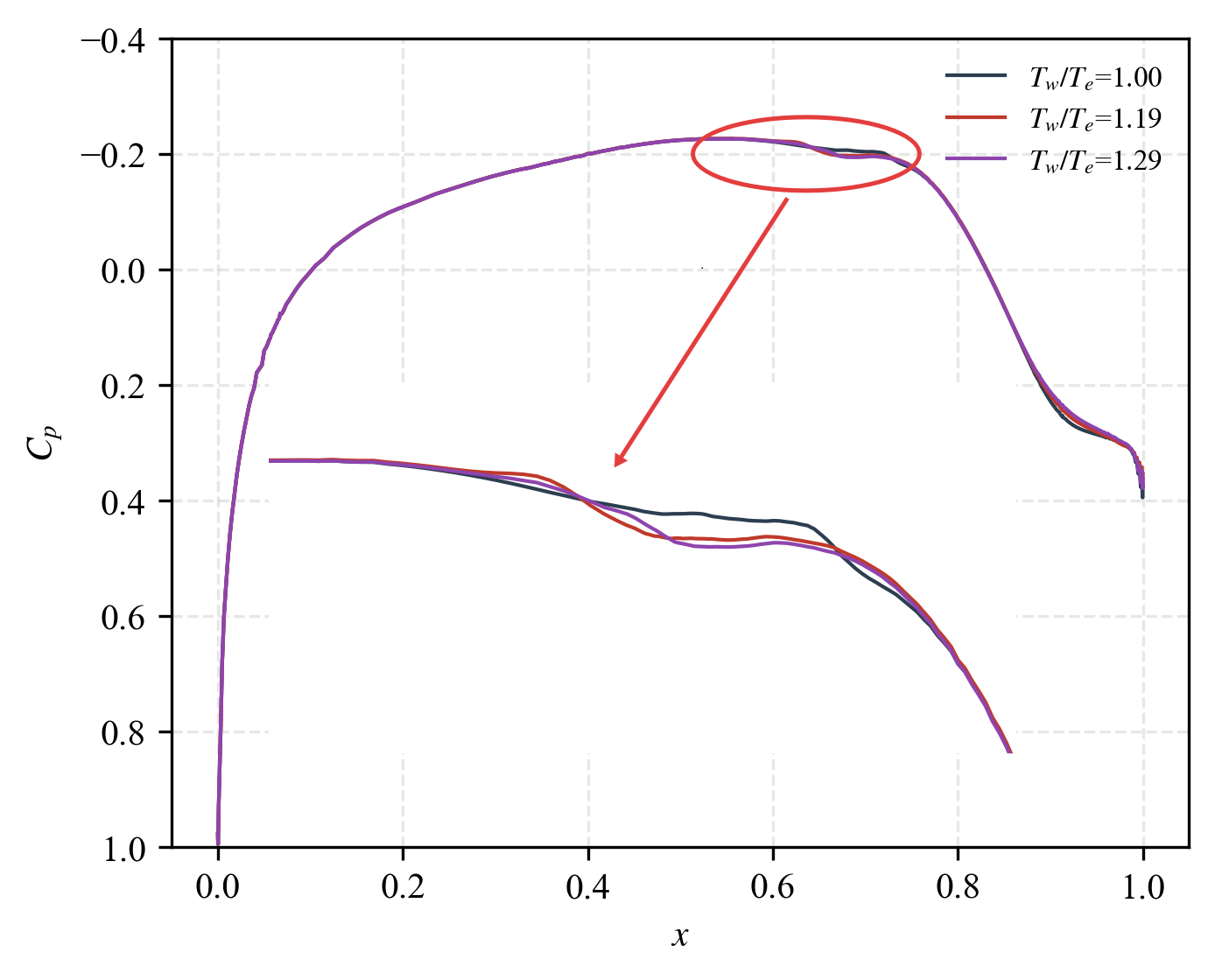}} 
	\subfigure[75$m/s$\label{f:Cp75}]
	{\includegraphics[height=2.2in]{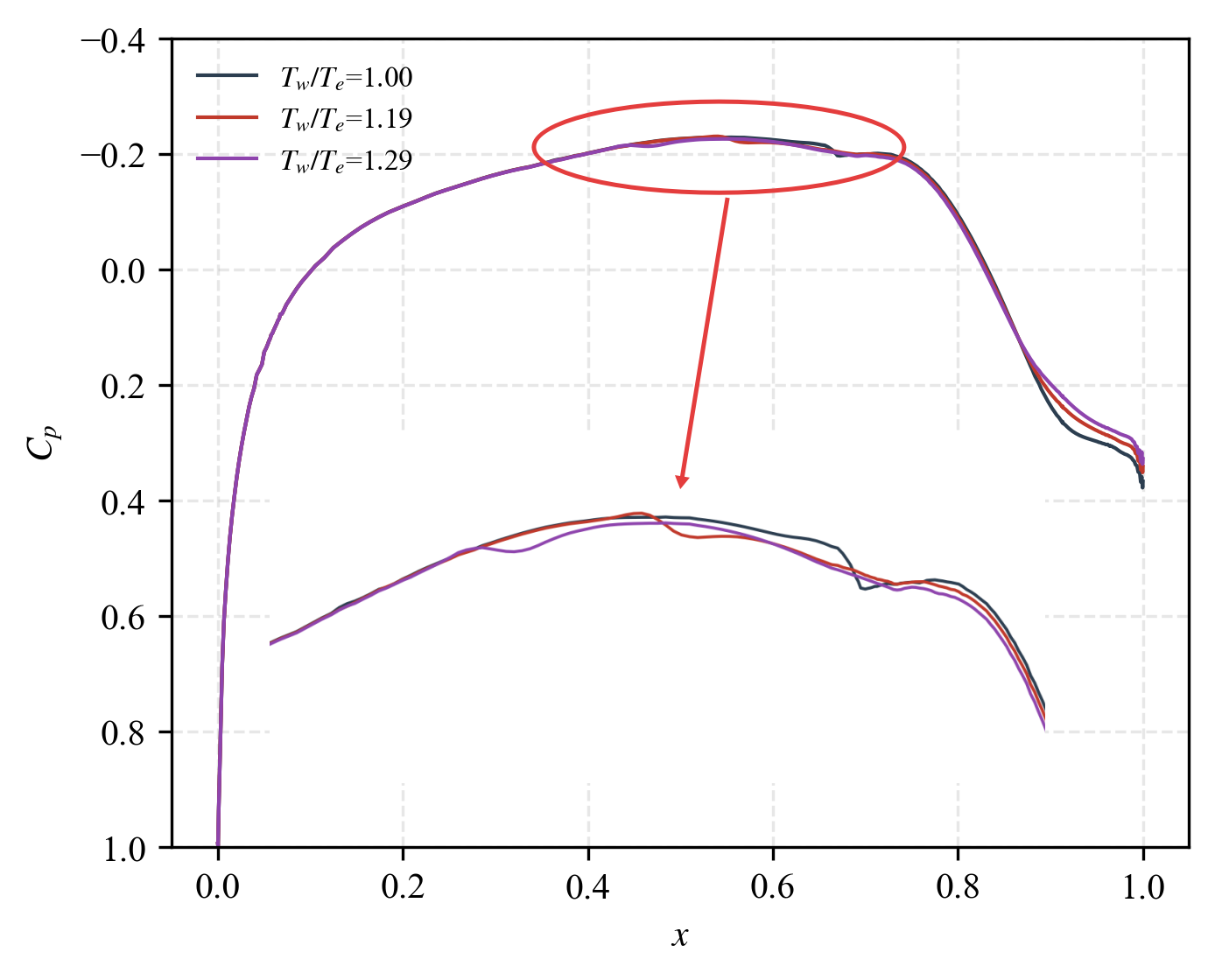}} 
	\caption{The pressure coefficient distribution along the streamwise direction for the airship.}
	\label{f:Cp}
\end{figure}

Here, we elucidate a critical phenomenon mentioned in Section \ref{s:WindTunnel}: at a velocity of 50 $m/s$, the transition location remains virtually unchanged with wall heating. However, at 75 $m/s$, the transition location shifts significantly under heating conditions. Fig.~\ref{f:Cp} illustrates the pressure distributions corresponding to different velocities. At 50 $m/s$, transition occurs in the adverse pressure gradient (APG) region, whereas at 75 $m/s$, due to the increased Reynolds number, transition occurs in the zero or weak favorable pressure gradient (FPG) region. As shown in Fig.~\ref{f:Retc_vs_lambda_theta}, in the APG region, wall heating has a minimal impact on the transition momentum thickness Reynolds number ($Re_{\theta t}$). Furthermore, the boundary layer in the APG region is thicker, requiring only a short streamwise distance to reach the critical $Re_{\theta t}$. In contrast, in the weak FPG or near-zero pressure gradient region, wall heating significantly impacts $Re_{\theta t}$. The corresponding boundary layer is thinner, necessitating a longer streamwise distance to attain the critical $Re_{\theta t}$. Thus, the combined effects of Reynolds number and local pressure gradient determine the influence of wall heating on the airship transition location.
Since the Reynolds numbers encountered in real airship flight are typically high, airships are therefore prone to surface-heating effects during daytime operation, which can degrade the drag-reduction performance. Consequently, the wall-heating effect must be considered.
With this improved transition model, we can proactively consider the adverse effects of daytime heating in future airship designs and develop aerodynamic shapes that can resist the shortening of the laminar region during the day, thus achieving robust laminar flow design.

\section{Conclusions}
\label{s:Conclusions}
Laminar drag reduction is an important means of reducing airship drag. However, during daytime operation, solar panels on the airship surface can raise the wall temperature far above the ambient temperature, leading to a wall-heating effect. Existing transport-based transition models have not yet accounted for this phenomenon. To address this issue, the main contributions of this work are as follows:
\begin{enumerate}
\item We conducted a series of linear stability theory (LST) analyses on FSC boundary layers under various wall-to-freestream temperature ratios, freestream turbulence intensities, and pressure gradient conditions. Based on the LST results, we developed a correction for the critical Reynolds number based on momentum thickness ($Re_\mathrm{\theta t}$) and the key parameter of ${Re_{v,\max}/Re_{\theta}}$ to account for wall-heating effects. Besides, we also consider the wall-cooling effects. The correction shows that wall heating reduces $Re_{\theta t}$, promoting earlier transition, while wall cooling increases $Re_{\theta t}$, delaying transition.
\item We integrated the improved $Re_{\theta t}$, ${Re_{v,\max}/Re_{\theta}}$ into the intermittency transport modeling~\cite{franccois2023simplified}. For the Schubauer and Klebanoff flat-plate case, we compared the improved model predictions with LST results under adiabatic, wall-cooling, and wall-heating conditions. The results demonstrate excellent agreement, with relative errors within 5\%.
Besides, the improved model and LST prediction agree well with the experimental result at the adiabatic condition.
This indicates that the improved model can accurately capture the effects of the wall-to-ambient temperature difference on transition.
\item Most importantly, we performed wind tunnel experiments on a heated airship model to investigate the effect of wall heating on transition location. The experimental results demonstrate that the impact of wall heating is governed by the combined effects of the Reynolds number and the local pressure gradient. Specifically, in the adverse pressure gradient region, the transition location is insensitive to wall heating; however, as the Reynolds number increases and the transition front moves into the favorable or zero pressure gradient region, wall heating significantly shifts the transition location upstream. We then validated the improved transition model against the experimental data, showing good agreement (relative errors no more 6\%) and confirming its effectiveness in capturing the influence of wall heating on transition.
\end{enumerate} 

Our improved stability-based transition model accounts for the effects of wall heating and wall cooling, providing an important transition-prediction approach to support future laminar-flow control for flight vehicles. In future work, we will extend the study to include crossflow instability and its dependence on wall cooling/heating temperature.

\section*{Acknowledgments}
This work was supported by the National Natural Science Foundation of China under grant number 12302302, 52572421 and the Taishan Scholars Program.

The authors acknowledge the assistance of AI for language polishing and improving the clarity of the manuscript.
\bibliography{mdolab-local}

% \section*{Appendix}
% The appendix is intentionally minimal in this condensed draft. Full experimental details, tables and figures can be restored on request.

\end{document}